\documentclass[usenatbib]{mnras}

\usepackage[T1]{fontenc}
\usepackage{ae,aecompl}

\usepackage{graphicx}	
\usepackage{amsmath}	
\usepackage{amssymb}	
\usepackage{bm}
\usepackage{ulem}
\usepackage{multirow}
\usepackage{balance}
\usepackage{times}

\newcommand{\mP}{{\mathcal{P}}}
\newcommand{\mC}{{\mathcal{C}}}

\newcommand{\dd}{\hbox{d}}

\newcommand{\rmh}{{\rm h}}
\newcommand{\rmm}{{\rm m}}

\usepackage{color}
\definecolor{Red}{rgb}{0.65,0.08,0.05}
\definecolor{Blue}{rgb}{0.05,0.08,0.65}
\definecolor{Purple}{RGB}{143,51,143}
\definecolor{Green}{rgb}{0.05,0.65,0.05}

\newcommand{\hGpc}[1]{\ensuremath{#1\, h^{-1}\,\mathrm{Gpc}}}

\title[Counts-in-cells statistics for biased tracers]{
A question of separation: disentangling tracer bias and \\
gravitational nonlinearity with counts-in-cells statistics
}

\author[C. Uhlemann, M. Feix, S. Codis, C. Pichon, F.~Bernardeau  et al.]{
C. Uhlemann$^{1}$\thanks{E-mail: c.uhlemann@uu.nl}, M. Feix$^{2}$, S. Codis$^{3}$, C. Pichon$^{2,4}$,  F.~Bernardeau$^{2,5}$,
\newauthor
\,  B. L'Huillier$^{6}$, J. Kim$^4$, S. E. Hong$^6$,  C. Laigle$^7$, C. Park$^4$, J. Shin$^4$ and D. Pogosyan$^8$\\
\\
$^{1}$ {Institute for Theoretical Physics, Utrecht University, Leuvenlaan 4, 3584 CE Utrecht, The Netherlands}\\
$^{2}$ {CNRS, UMR 7095 \& UPMC, Institut d'Astrophysique de Paris, 98 bis Boulevard Arago, 75014 Paris, France}\\
$^{3}$ {Canadian Institute for Theoretical Astrophysics, University of Toronto, 60 St. George Street, Toronto, ON M5S 3H8, Canada}\\
$^{4}$ Korea Institute of Advanced Study (KIAS) 85 Hoegiro, Dongdaemun-gu, Seoul, 02455, Republic of Korea\\
$^{5}$ {CNRS \& CEA, UMR 3681, Institut de Physique Th\'eorique, 91191 Gif-sur-Yvette, France} \\
$^{6}$ Korea Astronomy and Space Science Institute (KASI), 776 Daedeokdae-ro, Yuseong-gu, Daejeon 34055, Republic of Korea
\\
$^{7}$ Department of Physics, sub-department of Astrophysics, University of Oxford, Keble Road, Oxford OX1 3RH\\
$^{8}$ {Department of Physics, University of Alberta, 412 Avadh Bhatia Physics Laboratory, Edmonton, Alberta, T6G 2J1, Canada}
}

\date{Accepted XXX. Received YYY; in original form ZZZ}

\pubyear{2017}

\begin{document}
\label{firstpage}
\pagerange{\pageref{firstpage}--\pageref{lastpage}}
\maketitle

\begin{abstract}
Starting from a very accurate model for density-in-cells statistics of dark matter based on large deviation theory,  a bias model for the tracer density in spheres is formulated. It adopts a mean bias relation based on a quadratic bias model to relate the log-densities of dark matter to those of mass-weighted dark haloes in real and redshift space. The validity of the parametrised bias model is established using a parametrisation-independent extraction of the bias function.  This average bias model is then combined with the dark matter PDF, neglecting any scatter around it: it nevertheless yields an excellent model for densities-in-cells statistics of mass tracers that is parametrised in terms of the underlying dark matter variance and three bias parameters. The procedure is validated on measurements of both the one and two point statistics of subhalo densities in the state-of-the-art Horizon Run 4 simulation showing excellent agreement for measured dark matter variance and bias parameters. Finally, it is demonstrated that this formalism allows for a joint estimation of the nonlinear dark matter variance and the bias parameters using solely the statistics of subhaloes. Having verified that galaxy counts in hydrodynamical simulations sampled on a scale of 10 Mpc/h closely resemble those of subhaloes, this work provides  important steps towards making theoretical predictions for density-in-cells statistics applicable to  upcoming galaxy surveys like Euclid or WFIRST.
\end{abstract}



\section{Introduction}
\label{sec:intro}

Counts-in-cells statistics of galaxies have been extracted from observations in numerous works \citep{ShethMoSaslaw94,SzapudiMeiksinNichol96,Adelberger98,YangSaslaw11,Wolk13,Bel2016,Clerkin17,Hurtado-Gil17} spanning data sets from IRAS over SDSS to VIPERS and DES science verification. Conversely, significant theoretical progress has been made in analytically predicting the statistics of dark matter densities-in-spheres based on perturbation theory and local collapse models \citep[][providing only a non-exhaustive list of previous work]{Fry85,1989A&A...220....1B,Bernardeau92,Bernardeau94smooth,Bernardeau95,Juszkiewicz93,Juszkiewicz95,Munshi94,
ScoccimarroFrieman96,Fosalba98,Gaztanaga00,Valageas2002a,Ohta03}, which has been recently reformulated in terms of the theory of rare events \citep[][]{Bernardeau94rare,Valageas2002c,Bernardeau2014, Bernardeau2015,Bernardeau2016} with \cite{Uhlemann16log} achieving percent accuracy on the dark matter density PDF compared to state-of-the-art numerical simulations on scales of $\gtrsim 10$Mpc$/h$.

Such joint progress should now allow us to 
extract information from the mildly nonlinear regime 
so as to efficiently improve the estimation of cosmological parameters as this formalism allows for analytical predictions in this regime. Achieving this goal requires to relate the predictions for dark matter densities in spheres to  galaxy counts   which constitute biased tracers of the underlying matter field. 
Indeed, in addition to nonlinear gravitational dynamics and the effect of redshift-space distortions, clustering analyses of large-scale structure (LSS) are hampered by the fact that astronomical objects such as galaxies do not trivially trace
 the underlying dark matter distribution \citep[see][for a recent review]{Desjacques2016}. This problem has been known for a long time \citep[e.g.,][]{Abell1958, Dressler1980, Bahcall1983,
Kaiser1984, Coles1986}, and was subsequently confirmed in cosmological simulations demonstrating that haloes and galaxies are biased with respect to dark matter \citep[e.g.,][]{Cen1992, Kauffmann1997,
Blanton1999, Somerville2000}. Since then, several approaches have been pursued to accurately model these biasing relations. One main complication is that galaxy bias is generally a nonlocal and stochastic function of the dark matter field due to the varied physical processes partaking in galaxy formation \citep{Dekel1999, Scoccimarro2000}. 
Yet, smoothing the matter density fields over sufficiently large scales mitigates the effects of nonlocality and allows a sound description in terms of local bias expansions \citep[e.g.,][]{Fry1993} which aim at absorbing the underlying physics into a finite set of parameters. Later work has put such perturbative approaches onto firmer grounds by including nonlocal contributions and providing a consistent theoretical framework for the
statistics of biased LSS tracers \citep[e.g.,][]{Matsubara2011, Baldauf2011, Schmidt2013, Senatore2015, Porto2016}. Galaxies are believed to form inside the potential wells of dark matter haloes whose biasing properties can be systematically studied in numerical simulations or by means of analytic methods. Assuming that dark matter haloes are associated with peaks of the initial density field, the peak approach \citep{Kaiser1984, Bardeen1986} provides a nonperturbative model for biased populations and reasonably agrees with the abundance and the linear bias of
virialised haloes. Concerning nonlinearity as well as its dependence on other parameters like halo mass and scale, the bias of dark matter haloes is well approximated within the halo model \citep[e.g.,][]{Mo1996, Sheth1999, Cooray2002} based on the excursion set approach \citep{Bond1991}. Its relation to galaxies is typically quantified by combining cosmological N-body simulations with semianalytic models of galaxy formation \citep{Kauffmann1999, Berlind2002, Baugh2006, Mo2010}.

This paper will start from the dark matter side and make one crucial step towards reality by considering subhaloes, as the host of and proxies for galaxies and dark matter tracers. Such subhaloes can be extracted reliably from large cosmological simulations such as Horizon Run 4 \citep{HR4} that contain enough statistics to extract continuous PDFs. Note that the focus is on the issue of biasing for the PDF, such that it is in essence not so essential which tracers are chosen. However, the link between subhaloes and galaxies will also be discussed based on recent results from Horizon AGN \citep{Dubois14}, a cosmological hydrodynamical simulation that captures the evolutionary trends of observed galaxies over the lifetime of the Universe.  \cite{Bel2016} addresses the relation between continuous PDFs and discrete galaxy counts.

In general biasing is a notoriously challenging problem that requires the formulation of nonlocal and stochastic relationships between dark matter and tracer densities. This paper  will   however  show that for the purpose of obtaining the one- and two-point statistics of tracer densities in $\sim 10$Mpc$/h$ spheres, a mean local relationship (hence neglecting the scatter altogether) is enough to obtain predictions that are as accurate as the underlying statistics of dark matter densities. 
It will also show that the joint analysis of one- and two-cells counts allows us to lift the degeneracy between bias and dark matter variance,
  providing a key step towards making count-in-cells statistics applicable to  upcoming galaxy surveys like Euclid or LSST, for the purpose of 
  extracting cosmological parameters in the mildly non-linear regime.

This paper is organised as follows:  Section~\ref{sec:DMPDF} recaps the results presented in \cite{Uhlemann16log} for the dark matter density PDF.  Section~\ref{sec:halobias} turns to the bias between dark matter and tracer densities. After describing the Horizon Run 4 simulation and the halo identification scheme, an analytic bias model is formulated and compared to measurements from the simulation using scatter plots and a parametrisation-independent bias extraction. {Based on Horizon-AGN, the similarity of the mean bias relations for galaxies and halos is established and the influence of the scatter is assessed. Section~\ref{sec:HALOPDF} combines the bias model with the one-point dark matter PDF and two-point sphere bias to obtain the one-point halo PDF and two-point halo bias and establishes its accuracy against  simulations. }
 Section~\ref{sec:Applications} implements this formalism to estimate simultaneously  variance and biasing,
 and discusses applications and extensions. 
     Finally, Section~\ref{sec:Conclusion} concludes. 
     Appendix~\ref{sec:LogNormal}  compares the large deviation statistics (LDS) prediction to the  lognormal models.
Appendix~\ref{sec:PTdegen} shows perturbatively why the joint analysis of  the one- and two-point statistics breaks the degeneracy on tracer bias and dark matter variance.
Appendix~\ref{sec:hagn}  describes the hydrodynamical simulation Horizon-AGN.
     
\section{The dark matter density PDF}
\label{sec:DMPDF}

As  shown in \cite{Uhlemann16log}, the PDF for dark matter densities $\rho_\rmm$ within a sphere of radius $R$ at redshift $z$, valid in the mildly nonlinear regime, can be obtained from large deviation statistics (LDS) and is expressed as
\begin{equation}
\hskip -0.1cm \mP_R(\rho_\rmm) \!=\! \sqrt{\frac{\Psi''_{R}(\rho_\rmm)+\Psi'_{R}(\rho_\rmm)/\rho_\rmm}{2\pi \sigma^{2}_\mu}} \exp\left(-\frac{\Psi_R(\rho_\rmm)}{\sigma^{2}_\mu}\right ),
\label{eq:PDFfromPsi2}
\end{equation}
where the prime denotes a derivative with respect to $\rho_\rmm$ and
\begin{equation}
\Psi_R(\rho_\rmm)= \frac{\tau^{2}_{\rm SC}(\rho_\rmm) \sigma_L^{2}(R)}{2\sigma_L^2(R\rho_\rmm^{1/3})}\,.
\label{eq:Psiquad}
\end{equation}
Here $\sigma_\mu\equiv\sigma_\mu(R,z)$ is the nonlinear variance of the log-density (because the formula has been derived from an analytic approximation based on the log-density $\mu_\rmm=\log\rho_\rmm$) while $\sigma_L$ is the linear variance determined from the initial power spectrum $P_L$ using the Fourier transform of the spherical top-hat filter $W$
\begin{align}
\label{eq:linvariance}
\sigma_L(r)=\int \dd^3k\, (2\pi)^{-3} P_L(k) W(kr)^2\,.
\end{align}
$\tau_{\rm SC}(\rho_\rmm)$ is the linear density contrast averaged within the Lagrangian radius $r=R\rho_\rmm^{1/3}$ which can be mapped to the nonlinearly evolved density $\rho_\rmm$ within radius $R$ using the spherical collapse model. For this, an accurate approximation has been introduced by \cite{Bernardeau92} according to
\begin{equation}
\rho_{\rm SC}(\tau)=\left (1-\tau/\nu\right )^{-\nu} \, \Leftrightarrow\,
\tau_{\rm SC}(\rho)=\nu(1-\rho^{-1/\nu})\,,
\label{eq:spherical-collapse}
\end{equation}
where the parameter $\nu$ characterises the dynamics of spherical collapse. Here we choose $\nu=21/13$ to exactly match the high-redshift skewness obtained from perturbation theory
\citep{Bernardeau2014}. To ensure a unit mean density and the correct normalization of the PDF, one can simply evaluate the PDF obtained from equation~\eqref{eq:PDFfromPsi2} according to
\begin{align}
\label{eq:PDFfromPsi2norm}
\hat\mP_R(\rho_\rmm)= \mP_R\left(\rho_\rmm \, \frac{\langle\rho_\rmm\rangle}{\langle 1\rangle}\right) \cdot \frac{\langle\rho_\rmm\rangle}{\langle 1\rangle^2} \,,
\end{align}
with the shorthand notation $\langle f(\rho_\rmm)\rangle=\int_0^\infty \dd\rho_\rmm\, f(\rho_\rmm)\mP_{R}(\rho_\rmm) $. This step is necessary as equation~\eqref{eq:PDFfromPsi2} ensures the correct tree-level cumulants of order 3 and above, the right non-linear variance of $\mu_\rmm$ and zero mean for $\mu_\rmm$. If instead, one wants $\rho_\rmm$ to have unit mean, it is necessary to correct for the non-zero value of the mean of $\mu_\rmm$ using equation~\eqref{eq:PDFfromPsi2norm}.

Following \cite{Codis2016twopoint,Uhlemann17Kaiser},  the two-point PDF of the matter density reads in the large separation limit
\begin{equation}
\hskip -0.1cm
 \mP_{R}(\rho_\rmm,\rho_\rmm')\!=\! \mP_{R}(\rho_\rmm)\mP_{R}(\rho_\rmm')\! \left[1\!+\! \xi_{\circ,\rmm}(r) b_{\circ}(\rho_\rmm) b_{\circ}(\rho_\rmm') \right]\!, \label{eq:fullPDFlargeseparation}
\end{equation}
where $r>2R$ is the separation between two spheres of radius $R$ and densities $\rho_\rmm$ and $\rho_\rmm'$. The sphere bias encodes the excess correlation (with respect to the average sphere correlation $\xi_{\circ,\rmm}$) induced by a density $\rho_\rmm$ at separation $r$ and is defined as
\begin{equation}
\label{eq:def2ptbias}
b_{\circ}(\rho_\rmm)=\frac{\langle\rho_\rmm'|\rho_\rmm;r\rangle-1}{\xi_{\circ,\rmm}(r)}\ ,\ \xi_{\circ,\rmm}(r)=\langle\rho_\rmm\rho_\rmm';r\rangle-1\,.
\end{equation}
At large separation, it can be  computed with high accuracy using the large-deviation principle and is well approximated by
\begin{equation}
\label{eq:spherebias}
b_{\circ}(\rho_\rmm)=\frac{\tau_{\rm SC}(\rho_\rmm)\sigma_L^2(R)}{\sigma_L^2(R\rho_\rmm^{1/3})\sigma_\mu^2}\,,
\end{equation}
with once again a normalisation according to
\begin{equation}
\label{eq:spherebiasnorm}
\hat b_\circ(\rho_\rmm)=\frac{b_\circ(\rho_\rmm)-\langle b_\circ(\rho_\rmm)\rangle}{\langle (\rho_\rmm-1)b_\circ(\rho_\rmm)\rangle} \,.
\end{equation}

\section{Bias between matter and tracer densities}
\label{sec:halobias}
Let us now turn to biased tracers.  Section~\ref{subsec:Horizon} will first introduce the Horizon Run 4 simulation while  Section~\ref{subsec:bmodel} describes the theoretical models for tracer (galaxy and halo) bias.
  
\subsection{Biased tracers in Horizon Run 4 simulation}
\label{subsec:Horizon}

\subsubsection{Halo identification}

The Horizon Run 4 simulation \citep[HR4,][]{HR4} is a massive $N$-body simulation, 
evolving $6300^3$ particles in a \hGpc{3.15} box  using the GOTPM TreePM code \citep{GOTPM}. 
It assumes a WMAP-5 cosmology, with
$(\Omega_\text{m},\Omega_\Lambda,\Omega_\text{b},h,\sigma_8,n_\text{s}) = (0.26,0.74,0.044,0.72,0.79, 0.96
)$, yielding a particle mass of $9\times 10^{9} h^{-1}\,M_\odot$.
The initial conditions were generated at $z=100$ using the second order
Lagrangian perturbation theory, which ensures accurate power spectrum
and halo mass function at redshift 0 \citep{LHuillier14}.
The haloes were detected using Ordinary Parallel Friends-of-Friends
\citep[OPFOF,][]{KimPark06}, a massively parallel
implementation of the friends-of-friends 
(FoF) algorithm, using a canonical linking length of 0.2 mean particle
separations. Subhaloes were detected by the Physically Self-Bound algorithm
\citep[\textsc{psb},][]{KimPark06}, which finds the density peaks
within each FoF halo, removes  
unbound particles, similarly to the \textsc{subfind} halo finder, and additionally truncates the subhaloes to their
tidal radius.   All subhaloes with more than 30 particles were considered, yielding a masses from $2.7\times 10^{11}h ^{-1} M_\odot$ to $4.2 \times 10^{15}h ^{-1} M_\odot$.

\subsubsection{Weighting of halo densities}
Following the observations made in \cite{Jee2012} (Jee12 hereafter), let us consider a halo density with mass-weighting (instead of number-weighting) because this makes the bias relation much tighter and considerably reduces the scatter which is illustrated in Figure~\ref{fig:Scatter}. This observation can be understood by the intuition that mass-weighted halo densities resemble the overall dark matter density much more closely than halo number does. Note however that the mass-weighted densities of subhaloes are expected to be very similar to the mass weighted density of  haloes (with no substructure) as the mass is almost preserved from haloes to subhaloes. This paper considers subhaloes as defined in Section~\ref{subsec:Horizon} because they can be related to galaxies using abundance matching \citep{Kravtsov04,ValeOstriker04}, see Section~\ref{subsec:galaxies}. 

\subsection{Bias models: mean bias relations and their scatter}
\label{subsec:bmodel}
 \cite{Uhlemann16log} showed that the model for the PDF of the dark matter density field $\hat P_R(\rho_\rmm|\sigma_\mu)$ with the variance of the log-density $\sigma_\mu(R)$ as a driving parameter was accurate at the percent level for variances $\sigma\lesssim 0.5$. Hence, the question of how to obtain a similarly accurate model for the PDF of the density field of a biased mass tracer  
  boils down to successfully describing the effective bias relation between dark matter densities in spheres and the corresponding densities in spheres of their tracers. For simplicity,   this bias model  is formulated between dark matter and halo (or galaxy) densities for spheres of identical radii, so from now on $\rho_\rmm(\rho_\rmh)$ stands for $\rho_{m,R}(\rho_{h,R})$. While in general one would expect that the full joint PDF of dark matter and tracer densities is needed, including the scatter, it is shown in what follows that an accurate mean bias relation is enough to obtain an excellent model for the biased tracer PDF. This is in the spirit of large deviation statistics, that has been previously applied to argue that the mean local gravitational evolution given by spherical collapse is good enough to predict the dark matter PDF at fixed radius at percent accuracy\footnote{The large-deviation principle states that the statistics is dominated by the path that minimises the ``action'' -- or in our case the rate function -- in order to maximise the probability. This most likely path or dynamics can be decomposed into a gravitational part, given by the spherical collapse, and an astrophysical part, given by the mean bias relation.}.  
  
\subsubsection{Polynomial bias model in log-densities}
In order  to map the dark matter PDF to the halo PDF,  let us rely on an `inverse' bias model $\rho_\rmm(\rho_\rmh)$ writing the dark matter density as a function of the halo density which, according to Jee12,  has a better performance than the `forward' bias model $\rho_\rmh(\rho_\rmm)$. These bias parameters characterise the inverse relation and in particular our linear bias will typically have values around $1/2$ signalling positive linear forward bias around 2. Again, following Jee12, let us use a quadratic model for the log-densities $\mu=\log\rho$ (rather than for the densities) which reads
\begin{equation}
\mu_{\rm m} =  \sum_{n=0}^{n_{\rm max}} b_{n}\mu_{\rmh}^{n}\ ,\ n_{\rm max}=2\,.
\label{eq:POLYBIASloginv}
\end{equation}
It was checked that the higher order bias parameters are negligible, $|b_3|<0.002$ for all redshifts and radii considered here, and lead to very minor improvements of the quality of fit that do not warrant the use of this additional parameter. Note that, since the offset $b_0$ is additive in the log-densities, it ensures a multiplicative renormalisation for the density\footnote{When expanding the quadratic bias model for log-densities in the halo density contrast $\rho_\rmh=1+\delta_\rmh$ one obtains $$\rho_\rmm = \exp(b_0) \left( 1 + b_1 \delta_\rmh + \left[\frac{1}{2} (b_1-1) b_1+ b_2\right] \delta_\rmh^2 + \mathcal O(\delta_\rmh^3) \right)\,.$$ Interestingly, for the similar radii $R_1=10$, $R_2=15$Mpc$/h$ one finds identical $b_2$ and $\exp(b_0)b_1$ while $b_0$ and $b_1$ differ.}, {which is preferable according to an analytical result of \cite{FruscianteSheth12} that has been obtained from a lognormal mapping}.
Jee12  emphasize that  the reason why equation~\eqref{eq:POLYBIASloginv} can be approximated by a linear bias model for the density fluctuations $\delta_{\rmh} = \hat b_1\delta_{\rmm}$ on large scales is that the ranges of log-densities $\mu_\rmh$ and $\mu_\rmm$ become small and not because the bias relation itself becomes linear. This is particular relevant here when focussing  on the tails of the distribution of densities and hence the regime where linear bias is not sufficient.

\begin{figure}
\centering
\hskip -1cm \includegraphics[width=1.1\columnwidth]{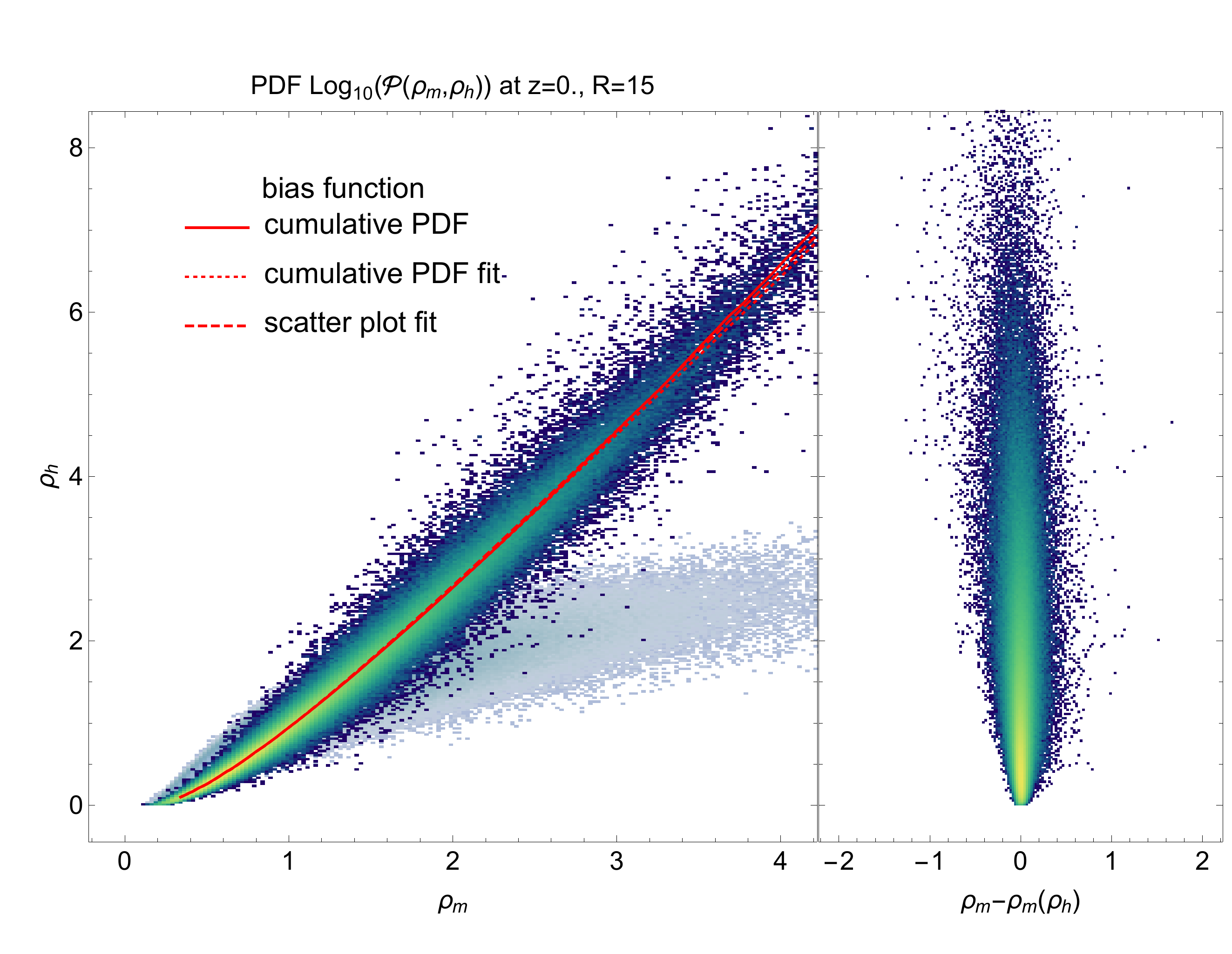}
\caption{(Left-hand panel) Density scatter plot of the halo density $\rho_{\rmh}$ with mass-weighting (blue-green region) and number-weighting (grey region) versus the dark matter density $\rho_{\rm m}$ for radius $R=15$ Mpc$/h$ at redshift $z=0$. The figure also shows the 
the best-fit quadratic bias model for the log-density obtained from a fit to the CDF bias function and the scatter plot (dotted and dashed line, respectively) which almost perfectly agrees with the parametrisation-independent bias obtained from the CDF (red line). Note that for mass-weighted halo densities, the scatter is reduced significantly compared to number-weighted halo densities.
(Right-hand panel) Residual scatter around the quadratic fit to the CDF bias function which is uniform and symmetric. }
\label{fig:Scatter}
\end{figure}

\begin{table}
\centering
\begin{tabular}{| ll | cccc | ccc}
\multicolumn{2}{c|}{param} & \multicolumn{2}{c|}{variance} & \multicolumn{2}{c|}{correlation} &\multicolumn{3}{c|}{bias}  \\\hline
$z$ & $R$ &  $\sigma_{\mu,\rmm} $& $\sigma_{\mu,\rmh} $& $\xi_{\rho,\rmm} $& $\xi_{\rho,\rmh} $ & $b_{0}$ & $ b_{1}$ & $ b_{2}$\\\hline
0 & 10 & 0.613 & 1.276 & 0.041 & 0.093 & 0.068 & 0.604 & 0.058 \\
0 & 15 & 0.475 & 0.855 & 0.043 & 0.099 & 0.036 & 0.618 & 0.058\\
1 & 10 & 0.411 & 1.006 & 0.015 & 0.067 & 0.054 & 0.460 & 0.055 \\
1 & 15 & 0.310 & 0.692 & 0.016 & 0.071 & 0.028 & 0.473 & 0.055 \\
\hline
$z$ & $R$ &  $\sigma_{\mu,\rmm}^z $& $\sigma_{\mu,\rmh}^z $& $\xi_{\rho,\rmm}^z $& $\xi_{\rho,\rmh}^z $ & $b_{0}^z$ & $ b_{1}^z$ & $ b_{2}^z$\\\hline
0 & 10 & 0.614 & 1.286 &  0.041 & 0.115 & 0.086 & 0.566 & 0.052\\
0 & 15 & 0.476 & 0.911 &  0.043 & 0.122 & 0.048 & 0.574 & 0.052\\
\end{tabular}
\caption{Collection of simulation results for different radii $R$ [Mpc$/h$] and redshifts $z$. The measured nonlinear variances $\sigma$ of the log-density $\mu=\log\rho$ and the correlation $\xi$ of the density $\rho$ at separation $r=30$Mpc$/h$ of both dark matter ($\rmm$) and haloes ($\rmh$) in real space (upper part) and redshift space (lower part) along with the bias parameters obtained from fitting the quadratic model from equation~\eqref{eq:POLYBIASloginv} to the bias function obtained from the CDF according to equation~\eqref{eq:CDFbias}. }
\label{tab:biasfit}
\end{table} 

\subsubsection{Parametrisation-independent inference of bias}
Following the idea of \cite{Sigad2000,Szapudi2004}, a direct way to obtain the mean bias relation is to use the properties of the cumulative distribution functions (CDFs), defined as $\mC(\rho)=\int_0^\rho d\rho' \mP(\rho')$,  so that
\begin{align}
\label{eq:CDFbias}
\mC_\rmm(\rho_\rmm)=\mC_\rmh(\rho_\rmh) \ \Rightarrow \ \rho_\rmm(\rho_\rmh)=\mC_\rmm^{-1}( \mC_\rmh(\rho_\rmh))\,.
\end{align}
This parametrisation-independent bias extraction is used to verify the accuracy of the polynomial log-bias model, equation~\eqref{eq:POLYBIASloginv}, as described below.

\subsubsection{Density scatter plots from numerical simulation}
Figure~\ref{fig:Scatter}   presents a  scatter plot showing $\rho_{\rm h}$ as a function of $\rho_{\rm m}$ for redshift $z=0$ and radius $R=15$ Mpc$/h$ in order  to assess how well bias models characterise the halo density bias. The lines correspond to the mean bias obtained in a parametrisation-independent way from the CDF method (red line) and fits based on a quadratic bias model for the log-densities (dotted and dashed red line) according to equation~\eqref{eq:POLYBIASloginv}. The corresponding values of the best-fit bias parameters are given in Table~\ref{tab:biasfit} for different redshifts and radii. The second-order bias model for the logarithmic densities based on equation~\eqref{eq:POLYBIASloginv} agrees almost perfectly with the parametrisation-independent way of inferring bias using CDFs as in equation~\eqref{eq:CDFbias} and matches simulation results very well, as has been observed in Jee12 for a wide range of mass cuts, smoothing lengths, and redshifts. Indeed, differences in the fits are almost imperceptible to the eye and at the sub-percent level throughout, except for the extreme low and high-density tail, and the residual scatter around the mean polynomial log-bias model is very symmetric and uniform. This has to be contrasted with a quadratic model in the mass-weighted halo densities that can be shown to have a clear residual skewness and to be significantly less accurate (residuals of about 2\% between $\rho\in [0.2,3]$, increasing more steeply in the tails). Since the mean bias relation is used to map the PDFs, having an even scatter around the mean relation is advantageous to mitigate possible effects of the scatter. Hence  in the following  the polynomial bias model for the log-densities will be used. 

\begin{figure}
\centering
\hskip -1cm\includegraphics[width=1.1\columnwidth]{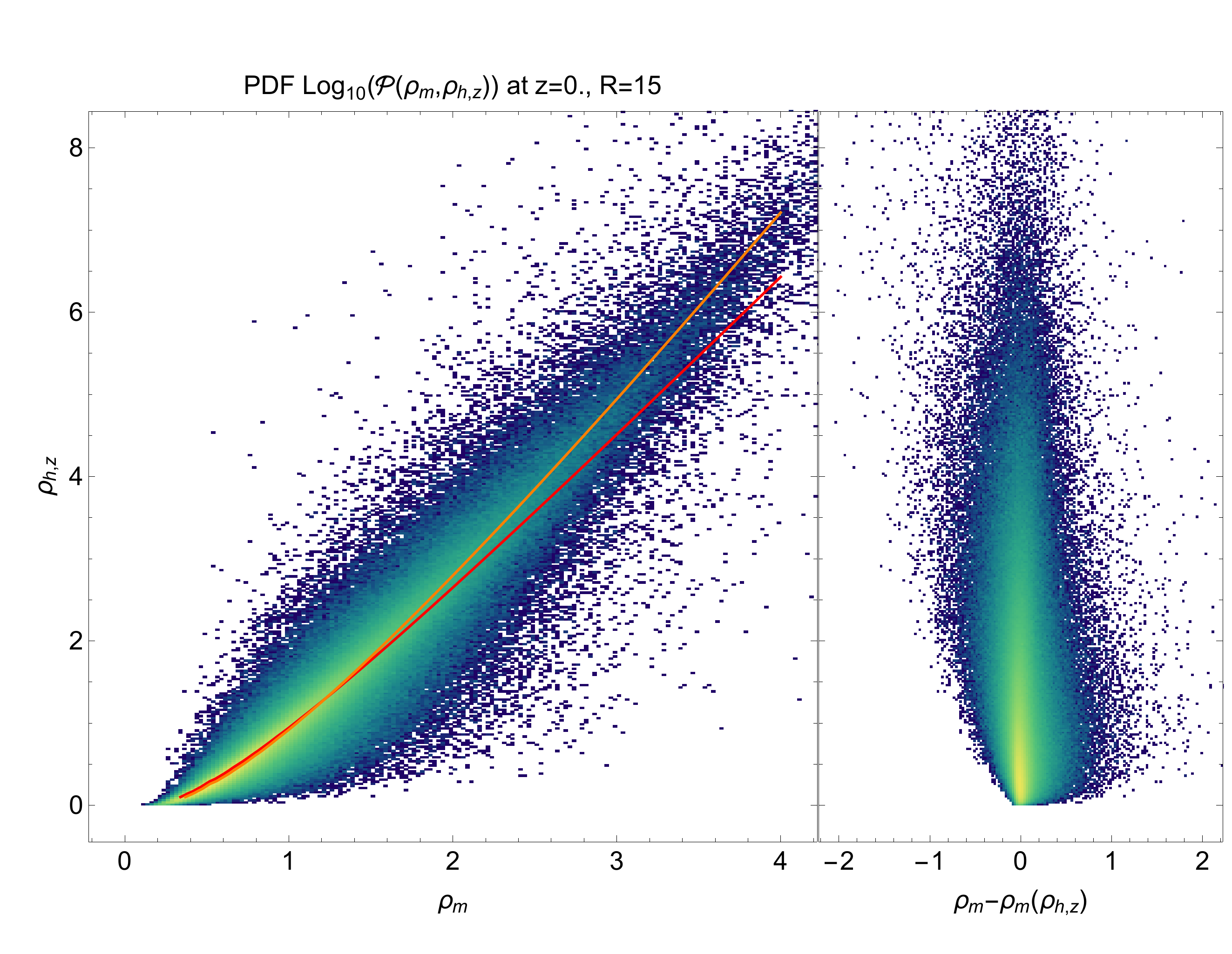}
\caption{(Left panel) Density scatter plot of the halo density $\rho_{\rmh,z}$ in redshift space with mass-weighting versus the dark matter density $\rho_{\rm m}$ for radius $R=15$ Mpc$/h$ at redshift $z=0$. The figure also shows the the best-fit quadratic bias model for the log-density obtained from a fit to the CDF bias function in real space (red line) and redshift space (orange line). 
(Right panel) Residual scatter around the quadratic fit to the CDF bias function which is uniform and symmetric (though the dispersion is somewhat larger 
than that of Figure~\ref{fig:Scatter}). }
\label{fig:Scatter_RSD}
\end{figure}

Furthermore, Figure~\ref{fig:Scatter_RSD} presents a scatter plot for the halo density determined in redshift space $\rho_{\rm h,z}$. As  was done in real space, a parametrisation-independent extraction of the mean bias relation was used  as a complement to the polynomial bias model in the log-densities \eqref{eq:POLYBIASloginv} for mass-weighted halo densities in redshift space, thereby extending the results of Jee12. When comparing the scatter plot from redshift space to its real space analogue (shown in Figure~\ref{fig:Scatter}), one can clearly see a enhanced scatter around the mean bias relation.  Yet, this extra scatter does not directly translate into inaccuracies of the PDF, as shown in Figure~\ref{fig:PDF_MEASURE_masslogbiasres}.

\subsubsection{Applicability to galaxies} 
\label{subsec:galaxies}
\begin{figure}
\includegraphics[width=1.\columnwidth]{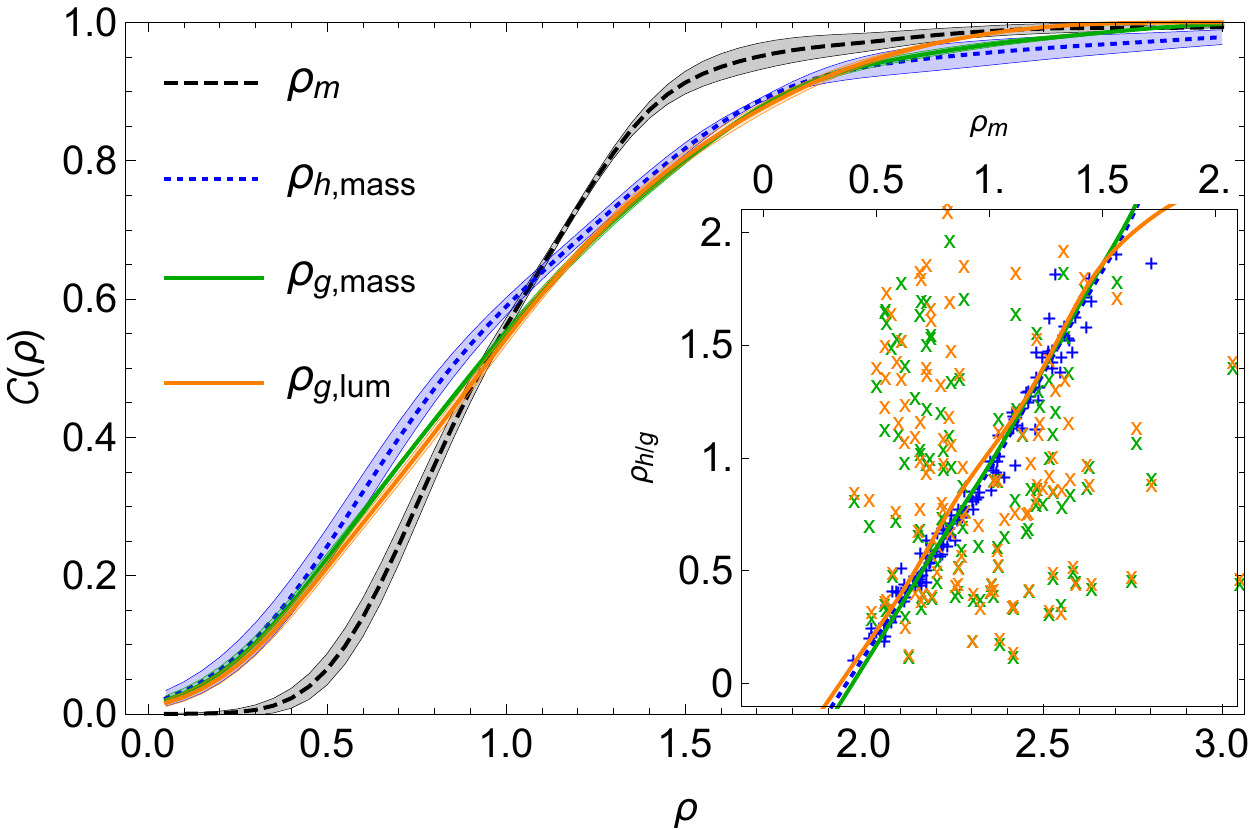}\\[-0.4cm]
\caption{The CDF for densities in spheres of radius $R=10$ Mpc$/h$ as measured  from Horizon-AGN at redshift $z=1$ for dark matter (dashed black line), mass-weighted subhaloes (dotted blue line), mass-weighted galaxies (solid green line) and luminosity-weighted galaxies (solid orange line). The shaded areas show an estimation of the error based on 5 subsamples. The inset  shows the scatter plot comparing mass- and luminosity-weighted galaxies (green and orange crosses) to mass-weighted subhaloes (blue plusses) including the bias function extracted from the CDF method from equation~\eqref{eq:CDFbias}.}
\label{fig:Scatter_halo_galaxies}
\end{figure}

In order to check to which extent our formalism developed for haloes will be applicable to galaxies, mass-weighted densities of haloes, galaxies and luminosity-weighted densities of galaxies were extracted from the state-of-the-art Horizon-AGN simulation, a full-physics hydrodynamical simulation in a cosmological volume \citep[][]{Dubois14}. Dark matter and mass-weighted subhalo densities in 125 non-overlapping spheres of radius $R=10$Mpc$/h$ are extracted from the simulated box at $z=1$. In order to mimic observational measurements, mass- and luminosity-weighted (in the $K_{\rm s}$-band) galaxy densities are extracted from the simulated lightcone  in a redshift range around $z=1$. Realistic galaxy luminosities have been computed in post-processing  using spectral synthesis, and galaxy stellar masses have been computed from photometry using SED-fitting, as usually done in observational datasets, which naturally allows to incorporate realistic errors (Laigle et al. in prep, see Appendix ~\ref{sec:hagn} for more details).
We didn't find any qualitative difference between the mean bias relations for galaxies and haloes. Indeed, Figure~\ref{fig:Scatter_halo_galaxies} displays the CDF of dark matter, mass-weighted subhaloes as well as mass- and luminosity-weighted galaxies together with the corresponding scatter plot. The blue, green and orange lines and points correspond to resp. mass-weighted subhaloes, galaxies and luminosity-weighted galaxies and are practically undistinguishable given the statistics we have\footnote{Note that, even for this state-of-the-art galaxy simulation, the box size is too small and hence the number of spheres not large enough to compare the PDFs directly.},  although the scatter of the galaxies is significantly increased compared to halos. This is a very promising result that motivates the use of mass-weighted halo density fields in this work. 
A thorough study of galaxy and halo bias in Horizon-AGN will be the topic of a forthcoming paper (Chisari et al, in prep.). 
{Note that, if one weights the galaxy densities with the mass of the host subhalo, the resemblance is even closer and the scatter reduced.} But in practice this would require both measuring the stellar masses (or luminosities) of the galaxies and relating them to the masses of the host subhaloes. The accuracy of the former is limited by the error on galaxy mass which is expected to be a function of the mass and redshift. At low redshift ($z<1$), the observed galaxy mass is generally underestimated compared to the intrinsic one and in general one can have a discrepancy up to $\Delta(\log M_{\rm g}) \simeq 15\%$ depending on the quality of the spectroscopy or photometry available to estimate the stellar mass \citep[see e.g.][Laigle et al. in prep]{Pforr12,Mobasher15}. {When adding a Gaussian noise of this size to the measured halo masses, as explicitly checked at $z=0$ for the radii $R=10, 15$ Mpc$/h$, the corresponding PDFs of the mass-weighted halo densities remain almost unchanged except for their deep tails. The best-fit bias parameters change only marginally, with the linear and largest bias parameter $b_1$ being most robust (sub-percent difference) and larger effects on the relatively small bias-renormalisation $b_0$ (5-7\% difference) and the quadratic bias $b_2$ (2-4\% difference).} For relating galaxy mass to halo mass, one can then use techniques based on subhalo abundance matching \citep[SHAM,][]{Behroozi10} or its extensions \cite[see e.g.][]{Yang12,KulierOstriker15}, which are very close in spirit to the modelling of bias used here and typically give an error of a similar size than the mass determination, at least for large halo masses. The same idea can be applied to galaxy luminosities \citep[see e.g.][]{ValeOstriker04,ValeOstriker06,Cooray05} which can be measured much more reliably than galaxy masses. Very recently, \cite{Moster17} presented an empirical model for galaxy formation 
finding that average star formation and accretion rates are in good agreement with models following an abundance matching strategy. One can also determine the galaxy-halo connection, in particular the stellar-to-halo mass ratio, from a joint lensing and clustering analysis of observations \citep[as done in][]{Coupon15,ZuMandelbaum15}
when using the halo occupation distribution (HOD) framework that assumes that the number of galaxies per halo is solely a function of halo mass, split into central and satellite contributions.

\section{The biased tracer density PDF}
\label{sec:HALOPDF}

Having established the accuracy of the bias model, let us now combine it with the one-point dark matter PDF and two-point sphere bias to obtain the one-point halo PDF and two-point halo bias. The accuracy of the analytical predictions for one- and two-point statistics will be checked against the simulation. In Appendix~\ref{sec:LogNormal} the analytical model for the halo PDF is compared to phenomenological reconstructions based on lognormal distributions and their extensions through cumulant expansions.   

\subsection{Mapping to the tracer PDF with the mean bias relation}

The halo density PDF, $\mP_{\rmh}$, can be generally written as a convolution of the dark matter PDF $\mP_\rmm$ and the conditional PDF of finding a certain halo density given a dark matter density 
\begin{equation}
\mP_{\rmh}\left (\rho_{\rmh}\right ) = \int\dd\rho_\rmm\, \mP_{\rm bias}\left(\rho_{\rmh}|\rho_\rmm\right)\mP_\rmm (\rho_\rmm),
\label{eq:CONVOLVE}
\end{equation}
where the conditional PDF $\mP_{\rm bias}(\rho_{\rmh}\vert\rho_\rmm)$ depends on the details of halo formation and its associated parameters such as, e.g., halo mass, smoothing scales, and redshift, but also includes stochasticity which results from an incomplete understanding of the formation process \citep[e.g.,][]{Dekel1999}. One could attempt to model the joint PDF with the help of simulated and observed datasets in the spirit of the halo model of galaxy clustering \citep[e.g.,][]{Cooray2002, Berlind2002}. Here, the scatter around the mean relation between $\rho_\rmm$ and $\rho_{\rmh}$ will be neglected: this nonetheless leads to an excellent model for the halo PDF provided the underlying bias model is appropriate. Equipped with a bias model for the mean relation $\rho_\rmm(\rho_{\rmh})$, the halo PDF $\mP_{\rmh}$ is now obtained from the dark matter PDF $\mP_\rmm$ in equation~\eqref{eq:PDFfromPsi2} by conservation of probability
\begin{equation}
\mP_{\rmh}\left (\rho_{\rmh}\right ) = \mP_\rmm(\rho_\rmm (\rho_{\rmh})) \left\lvert \dd\rho_\rmm/\dd\rho_{\rmh}\right\rvert\,,
\label{eq:HALOPDF}
\end{equation}
where it is assumed that $\rho_\rmm(\rho_{\rmh})$ is a strictly monotonic function. 
Using equation~\eqref{eq:fullPDFlargeseparation},  the halo two-point PDF  can eventually be written down as
\begin{align}
\label{eq:fullPDFlargeseparationhalo}
\mP_\rmh(\rho_\rmh,\rho_\rmh')=& \mP_\rmh(\rho_\rmh)\mP_\rmh(\rho_\rmh') \times \notag\\
&\left[1+ \xi_{\circ,\rmm}(r) b_{\circ,\rmm}(\rho_\rmm(\rho_\rmh)) b_{\circ,\rmm}(\rho_\rmm'(\rho_\rmh')) \right]\,, 
\end{align}
One can then define the modulation of the two-point correlation function, the sphere bias $b_\circ$ for halos from the result for dark matter given in equation~\eqref{eq:spherebias}
\begin{equation}
\label{eq:spherebiashalo}
b_{\circ,\rmh}(\rho_\rmh) = b_{\circ,\rmm}\left(\rho_\rmm(\rho_\rmh)\right) \sqrt{\xi_{\circ,\rmm}/\xi_{\circ,\rmh}} \,,
\end{equation}
where the ratio of correlation functions is given by
\begin{equation}
\sqrt{\xi_{\circ,\rmh}/\xi_{\circ,\rmm}}=\left\langle \rho_{\rmh}(\rho_{\rmm}) b_{\circ,\rmm}(\rho_{\rmm})\right\rangle,
\end{equation}
and can be approximated by expanding the log-bias relation to first order to obtain $\sqrt{\xi_{\circ,\rmm}/\xi_{\circ,\rmh}}
 \simeq \exp(b_0) b_1$.

\subsection{Checking the accuracy of  halo PDF against simulations}
\begin{figure}
\centering
\includegraphics[width=\columnwidth]{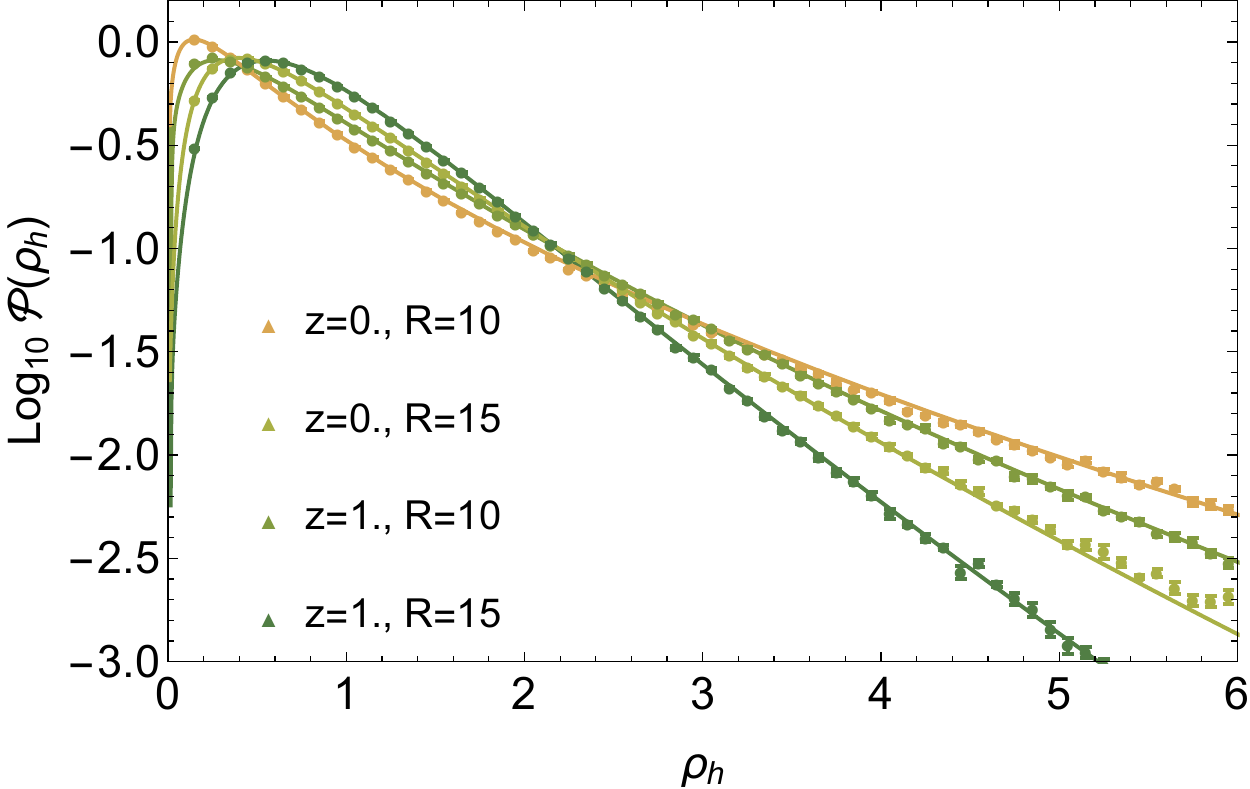}\\[-0.2cm]
\caption{Logarithmic view on the halo PDF as measured from the simulation in real space (data points) at redshifts $z=0,1$ for radii $R=10, 15$ Mpc$/h$ and analytically predicted (lines) using the measured dark matter variance and bias parameters given in the upper part of Table~\ref{tab:biasfit}.}
\label{fig:PDF_MEASURE_masslogbias}
\end{figure}

\begin{figure}
\hskip -0.5cm\includegraphics[width=1.1\columnwidth]{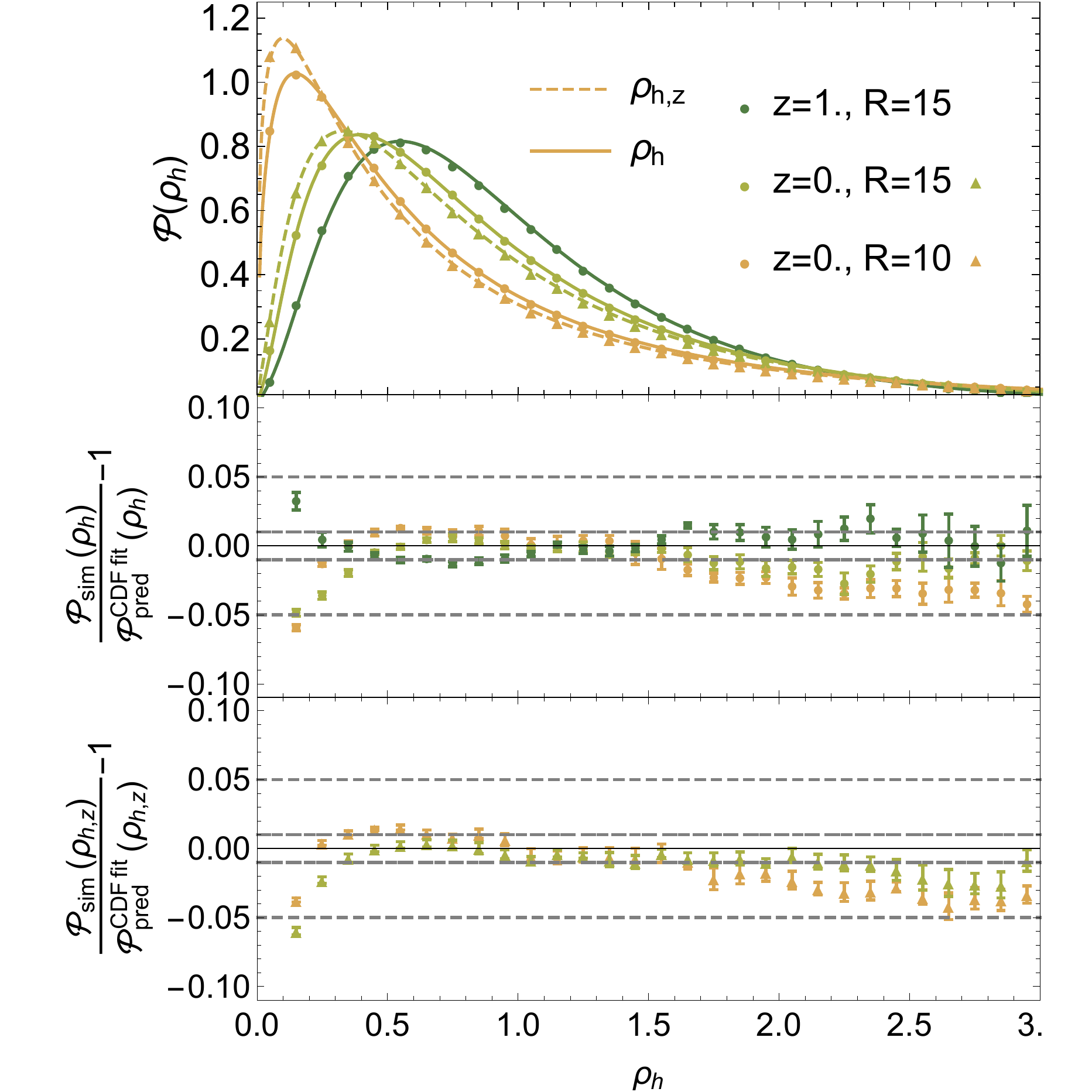}\\[-0.6cm]
\caption{(Top panel) Halo mass-density PDFs $\mP_{\rmh}$ for measurements based on halo catalogues in real space (points) and redshift space (triangles). Shown are results for the quadratic bias for the log-densities models in real space (solid lines) and redshift space (dashed lines) with fit values according Table~\ref{tab:biasfit}. (Middle and bottom panel) The corresponding residuals in real space (middle) and redshift space (bottom).}
\label{fig:PDF_MEASURE_masslogbiasres}
\end{figure}

\begin{figure}
\hskip -0.5cm\includegraphics[width=1.1\columnwidth]{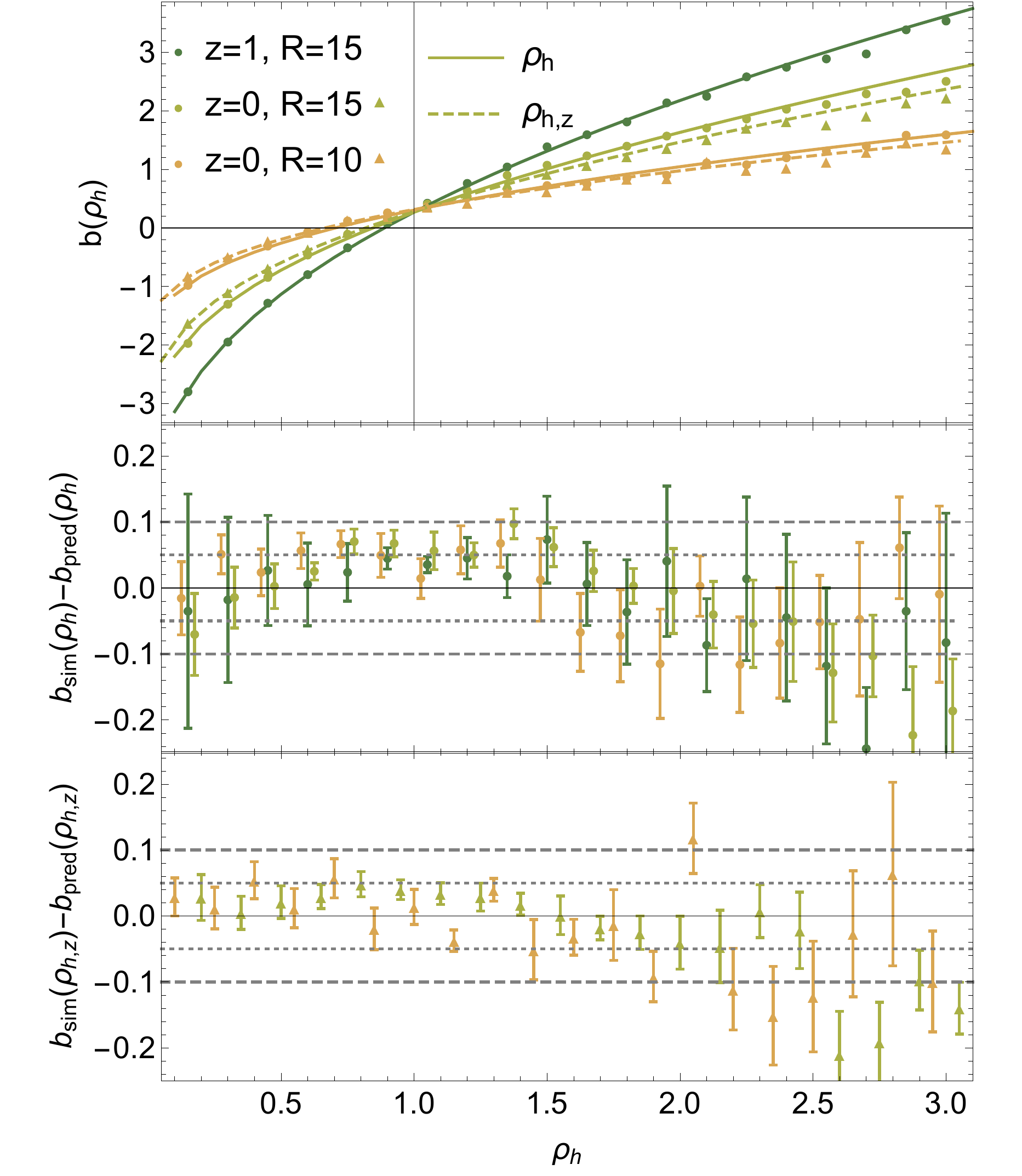}\\
\caption{Halo sphere bias function $b_\circ$ describing the modulation of the two-point halo correlation function with the density as measured for a separation $d=30$Mpc$/h$ in real space (circles) and redshift space (triangles) in comparison to the analytical prediction based on a measured dark matter variance $\sigma_\mu$ and the fitted bias parameters in real space $b_n$ (solid lines) and redshift space $b_n^z$ (dashed lines) as given in Table~\ref{tab:biasfit}.}
\label{fig:2ptbiashalo}
\end{figure}

 Figure~\ref{fig:PDF_MEASURE_masslogbias}~and~\ref{fig:PDF_MEASURE_masslogbiasres}  show the result of the halo-PDF obtained from \eqref{eq:HALOPDF} using the measured variance of the dark matter log-density and the best-fit bias parameters for the bias model for the log-densities up to second order reported in Table~\ref{tab:biasfit}. The  prediction for the halo PDF clearly matches the data, presenting residuals at the percent level in a wide range of halo densities from 0.2 to 3, in both real and redshift space. This should be contrasted to the log-normal PDF family discussed in Appendix~\ref{sec:LogNormal}.
 This is very encouraging given the  level of non-linearities involved in halo formation. The scatter of the bias relation could in principle have degraded the accuracy of the PDF, but Figure~\ref{fig:PDF_MEASURE_masslogbiasres} shows that it turns out to be a small effect. This remains true for counts of halos in redshift space, even though the redshift space scatter plot displayed significantly larger scatter than its real space counterpart.
Figure~\ref{fig:2ptbiashalo} compares the prediction for the sphere bias function in both real and redshift space, based on the same inputs as used for the halo PDF, with the measurements from the simulation and is also displaying excellent agreement. {Note that, for the redshift-space correlation, which has an angular dependence,  only   the monopole is effectively probed.}
To measure the sphere bias function, encoding the excess correlation between densities in spheres according to equation~\eqref{eq:def2ptbias}, a separation of $r=30$Mpc$/h$ is chosen, giving a grid of non-overlapping spheres. {The densities of the 6 neighbouring spheres} are collected in bins of width $\Delta\rho=0.15$; precise formulas are given by equations (19) and (20) in \cite{Uhlemann17Kaiser}.

\section{Application: parameter estimation}
\label{sec:Applications}

One of the main goals of constructing tracer statistics is to extract cosmological parameters from counts-in-cells.
Let us now make use of  the one-point halo PDF~\eqref{eq:HALOPDF} alone or combine it with the density-dependent sphere bias~\eqref{eq:spherebiashalo} to estimate either the bias parameters, the underlying dark matter variance, or both. 

Due to the strong (although not complete) degeneracy between the dark matter variance and linear bias (that can be shown to hold exactly for a linearly biased lognormal PDF (see Appendix~\ref{sec:LogNormal}), {and is given as $\sigma_{\mu,\rmh}\simeq\sigma_{\mu,\rmm}/b_1$ at leading order in perturbation theory according to equation~\eqref{eq:b1deg}}), it turns out one cannot use the one-point statistics alone to jointly determine the dark matter variance and bias parameters. This is not at all surprising, given the well-known degeneracy between linear bias and the clustering amplitude, caused by the fact that a low matter fluctuation amplitude can be masked out by a high galaxy bias or vice versa \citep[see e.g.][]{Seljak05}. In principle, if (i) all the statistics could be measured exactly,
(ii) the truncation in the bias model was fully justified, and (iii) the dark matter PDF was exactly given by the LDS model and in particular different from log-normal, 
then it should be possible to measure jointly the dark matter variance and the three bias parameters. In practice, when considering limited noisy samples, only the first three cumulants (mean, variance, skewness) carry enough information in a statistical sense, so that measuring the one-point PDF can only put three constraints on the parameters of the model. For a  quadratic bias model, this means that one effectively  ends up with a degeneracy line (i.e a one-dimensional manifold) in the four dimensional parameter space. 
Indeed, subsection~\ref{sec:oneptapplication} shows how in practice the one-point model does not yield enough information to measure both on realistic surveys  and discusses complementary strategies when relying on one-point statistics only, 
while
  subsection~\ref{sec:basic} explains why one- and two-points halo counts does  break this degeneracy in principle.
Finally subsection~\ref{sec:twoptapplication}  shows   how   a joint fit of both   counts from the  HR4 simulation yields an estimate of all four parameters plus the dark matter correlation function. 

\subsection{Bias-variance degeneracy in one-point statistics}
\label{sec:oneptapplication}

\begin{figure}
\centering
\includegraphics[width=1\columnwidth]{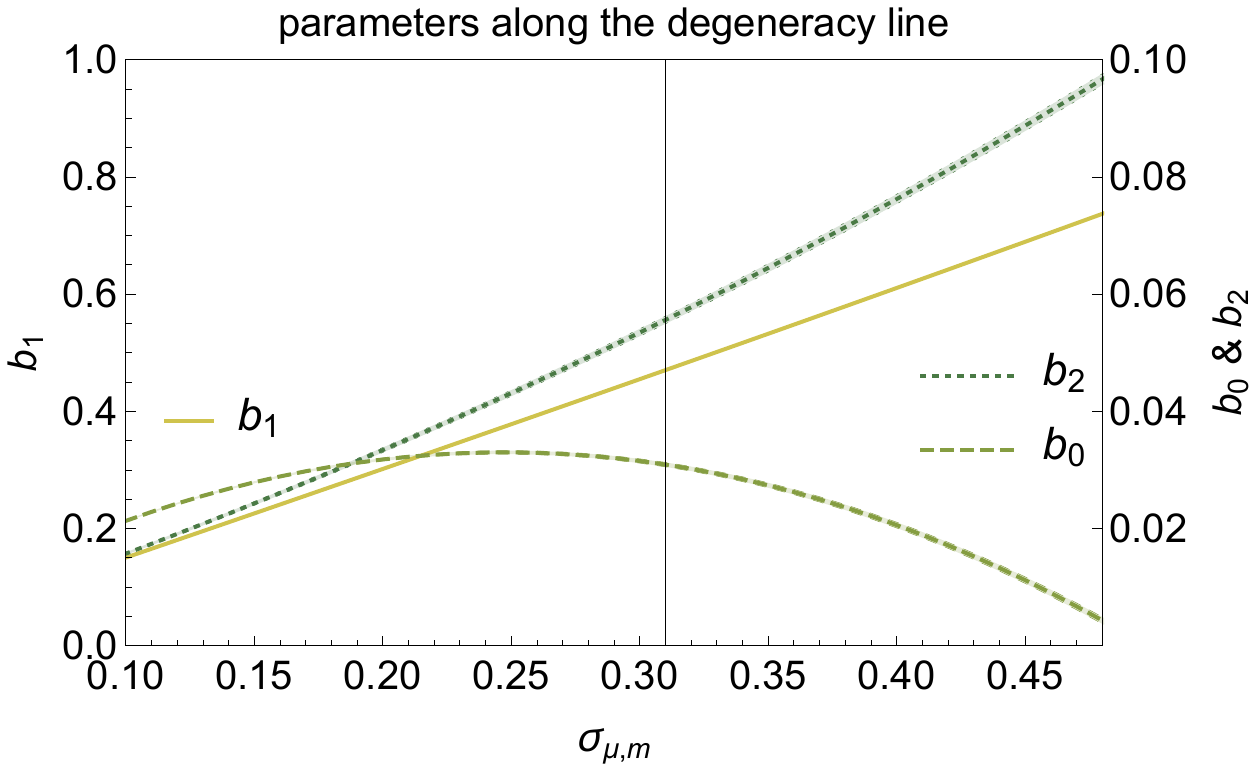}\\
\hspace{-0.8cm}
\includegraphics[width=0.96\columnwidth]{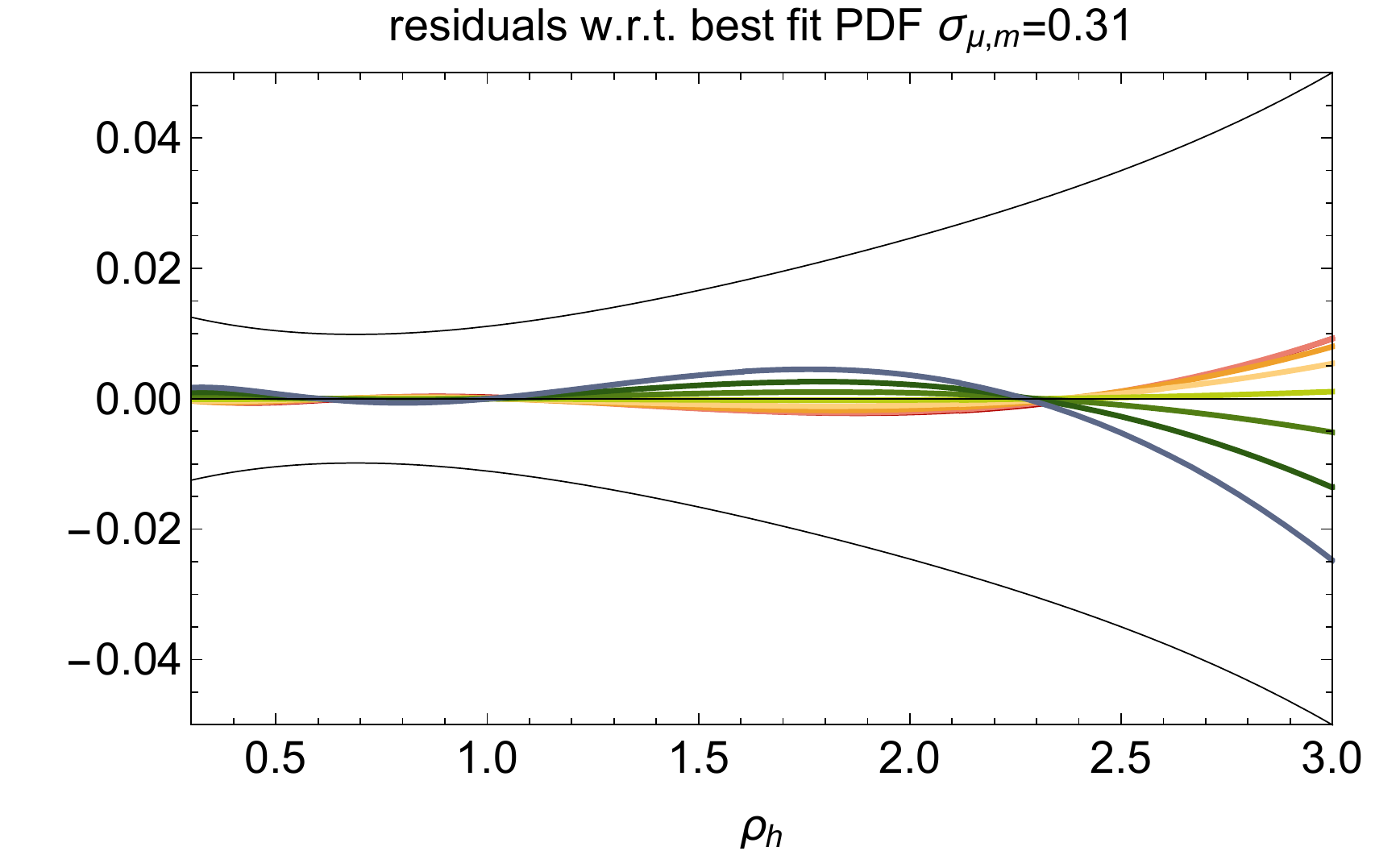}\\
\caption{Top: Parameters along the degeneracy line obtained from a fit to the measured density PDF at $z=1$ and for a radius $R=15$Mpc$/h$ in the Horizon Run 4 simulation when determining the bias parameters $b_n$ given a fixed dark matter variance $\sigma_{\mu,\rmm}$. The thin shaded area corresponds to the one-sigma confidence interval for different values of the $\sigma_{\mu,\rmm}$. 
Bottom: predicted density PDF along the degeneracy line from $\sigma_{\mu,\rmm}=0.1$ (red) to 0.45 (blue). Only residuals compared to the true value $\sigma_{\mu,\rmm}=0.31$ are displayed. The black lines show the one-sigma error on the measured PDF, obtained by fitting with a polynomial the binned one-sigma error bars.}
\label{fig:dline}
\end{figure}

In order to quantify the bias-variance degeneracy in one-point statistics, let us measure  the density PDF  at $z=1$ in the Horizon-run 4 simulation covered by spheres of radius $R=15$Mpc$/h$,  and get one-sigma error bars as the error on the mean estimated from 8 subcubes. Let us describe the degeneracy with $\sigma_{\mu,\rmm}$ as the curvilinear coordinate and for each value of $\sigma_{\mu,\rmm}$ between 0.1 and 0.5, and fit the measured non-linear PDF from $\rho_{\rmh}=0.3$ to $3$ with bins $\Delta\rho_{\rmh,\mP}=0.01$ as this is the regime where the model is expected to work well. The one-sigma confidence intervals of the bias parameters as a function of $\sigma_{\mu,\rmm}$ are displayed in the top panel of Figure~\ref{fig:dline}. As expected from the perturbative argument, the degeneracy line is dominated by a linear relationship between $b_{1}$ and $\sigma_{\mu,\rmm}$ (with slope $\sigma_{\mu,\rmh}\approx0.7$) with higher order correction leading to non-zero (but small) values of $b_{0}$ and $b_{2}$. {The parabolic shape of $b_0$ and the linear growth of $b_2$ with $\sigma$, as well as their smallness, can in fact be understood perturbatively, as shown in equations~\eqref{eq:b0ofb1b2} and \eqref{eq:b2deg} in Appendix~\ref{app:2ptbias}.} 
The bottom panel of Figure~\ref{fig:dline} shows that the predicted PDFs along the degeneracy line are all within the one-sigma error bars of the simulation and therefore cannot be distinguished.  
Combining this observable with other probes or using a model for the dark matter variance should in principle break this degeneracy.

 \begin{figure*}
\includegraphics[width=2\columnwidth]{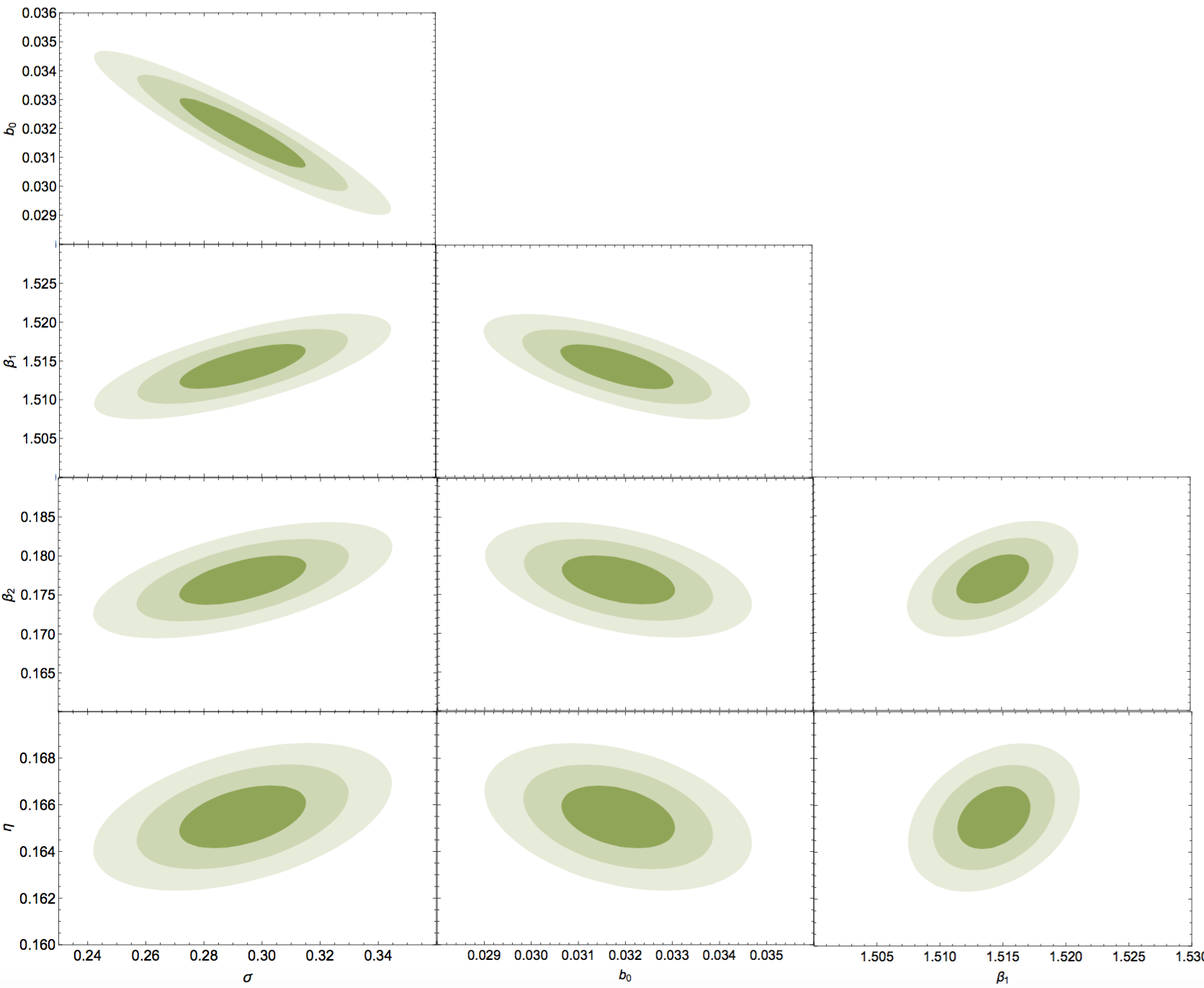}
\caption{One, two and three sigmas contours obtained by fitting the density PDF and the bias function at $z=1$ and for spheres of radius $R=15$Mpc$/h$ where $\eta=\xi/\sigma^{2}$, $\beta_{i}=b_{i}/\sigma$ and $\sigma=\sigma_{\mu,\rmm}$.}
\label{fig:FoM-full}
\end{figure*}

If the nonlinear dark matter variance was known, for example from empirical relations found in simulations \citep[such as][]{ReppSzapudi17} or higher order perturbation theory \citep[see e.g.][]{ScoccimarroFrieman96}, one could use the analytic dark matter PDF \eqref{eq:PDFfromPsi2} to obtain the CDF of dark matter $\mathcal C_\rmm$ and then the bias relation using equation~\eqref{eq:CDFbias} by measuring the halo CDF $\hat{\mathcal C}_\rmh$. Note that this procedure essentially looks for a nonlinear transformation of halo-densities such that the result is distributed according to the dark matter PDF equation~\eqref{eq:PDFfromPsi2}, and hence similar in spirit to the idea of Gaussianising the field \citep[see e.g.][]{McCullagh16}.

Conversely, if the bias parameters (including their time evolution) were known from either theory or measured from an independent probe, one could use the analytic halo PDF~\eqref{eq:HALOPDF} to determine the dark matter variance and use this to constrain for example the dark energy equation of state as demonstrated for dark matter in \cite{Codis2016DE}. Analytical attempts to predict cumulants of the halo density have been based on bias models starting from Press-Schechter \citep{Casas-Miranda02,Casas-Miranda03}, its extensions like excursion sets or peak theory, or the halo model \citep{Fry11}. {Note that, to take advantage of this idea one needs access to the bias that relates averaged halo and matter densities rather than the bias based on n-point functions. While there is a mapping between the two in the large-scale limit, for $R\gtrsim 50$Mpc$/h$, they are not equal and their relation depends on the shape of the power spectrum as well as the smoothing radius and filter shape, as pointed out in \cite{Desjacques2016}.} For particular observational signatures that are not degenerate with bias, such as local primordial non-Gaussianity (Uhlemann et. al. in preparation), the present formalism allows to take the nature of tracers into account and hence to obtain more realistic constraints. In principle, future peculiar velocity surveys could also gain us qualitative insights into biasing following the idea described in \cite{Uhlemann16vel}, although their statistical power is unlikely to yield accurate enough constraints.

\subsection{Joint one- and two-point statistics: the basic idea} 
\label{sec:basic}

\begin{figure}
\includegraphics[width=1.0\columnwidth]{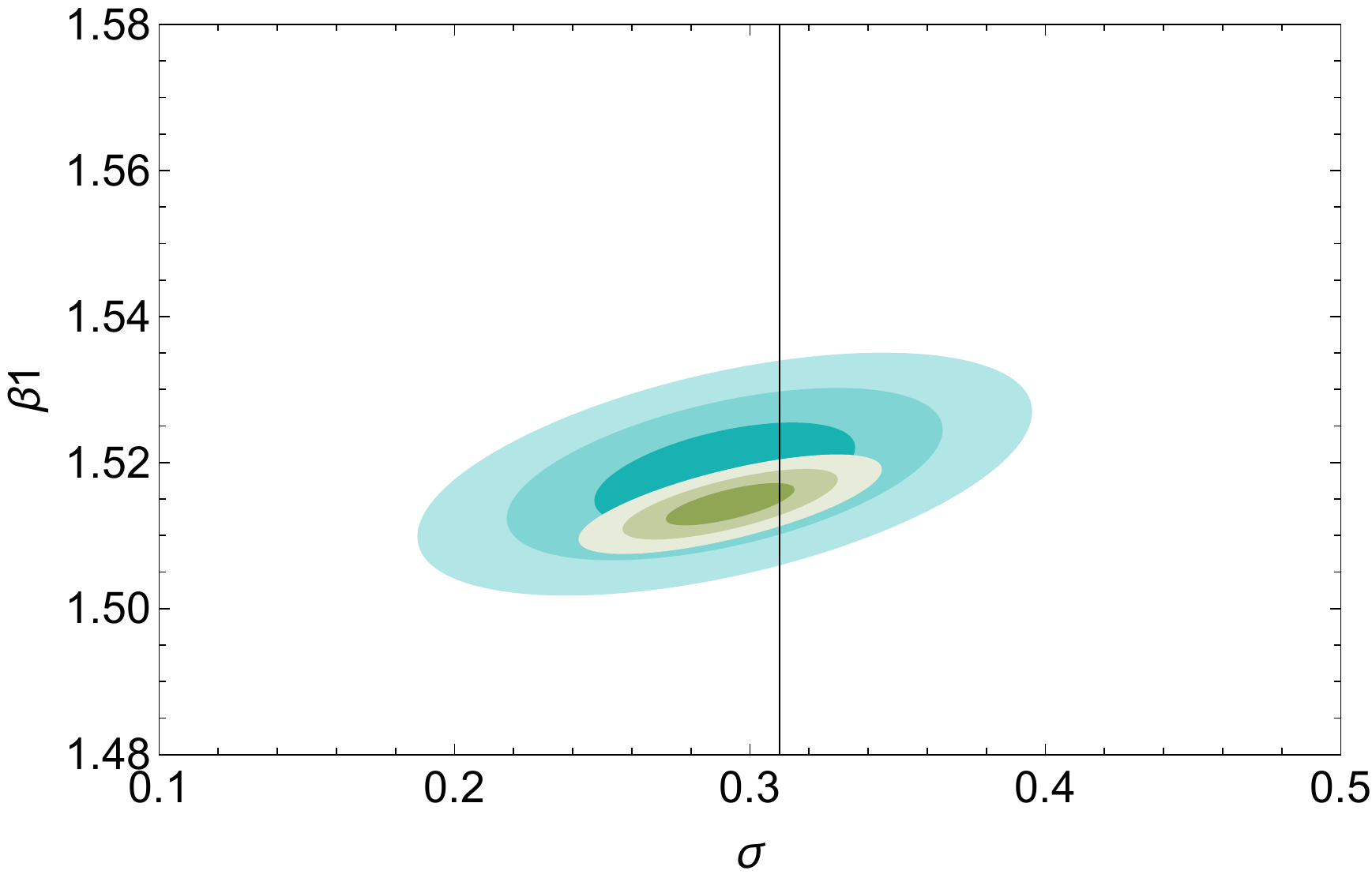}
\caption{
Mean one, two, three sigmas contours obtained from the 8 subcubes by averaging the best fits and covariance matrix (cyan). For comparison, the figure of merit of the whole volume is superimposed in dark green and a line at the target value $\sigma\equiv\sigma_{\mu,\rmm}=0.31$ is displayed. As expected the constraints on the model parameters shrink when the accessible volume increases.
\label{fig:subsampling}
}
\end{figure}

In order to break the degeneracy between bias parameters and the dark matter variance, one can make use of the two-point statistics from equation~\eqref{eq:fullPDFlargeseparationhalo} to jointly constrain the dark matter variance and biases. The  two-point halo PDF is built from the one-point halo PDFs~\eqref{eq:HALOPDF} and the density-dependent sphere bias~\eqref{eq:spherebiashalo} that modulates the two-point correlation function which were successfully compared to numerical simulations  in Section~\ref{sec:HALOPDF}.

Let us present here the basic idea behind the degeneracy lift.
The leading-order mixed cumulant depends on the two-point sphere bias function via 
\begin{equation}
\label{eq:defC12}
C_{12,\rmh}=\langle \delta_\rmh^2{\delta_\rmh'}\rangle = \xi_{\circ,\rmh} \int (\rho_\rmh-1)^2 b_{\circ,\rmh}(\rho_\rmh) P(\rho_\rmh)\, \dd\rho_\rmh\,.
\end{equation}
 Since the sphere bias function is not linear $b_{\circ,\rmh}(\rho) \nsim \rho_\rmh-1$, especially in the tails that are sensitive to $b_2$, 
equation~\eqref{eq:defC12}
 differs from  the  one-point cumulant given by the skewness
\begin{equation}
S_{3,\rmh}=\langle \delta_\rmh^3\rangle = \int (\rho_\rmh-1)^3 P(\rho_\rmh)\, \dd\rho_\rmh\,.
\end{equation}
The leading order expressions\footnote{Note that, at that order, the cumulants of the density $\rho$ and log-density $\mu$ only differ by a constant, see \cite{Uhlemann16log}. 
} relating the corresponding dark matter and halo cumulants for the adopted (inverse quadratic in the log-densities) biasing model are consistently given by
\begin{align}
\label{eq:S3bias}
S_3^{\mu,\rmm}=S_{3}-3 &=b_1^{-1} \left(S_3^{\mu,\rmh}+6 b_2/b_1\right)\,,\\
\label{eq:C12bias}
C_{12}^{\mu,\rmm} =C_{12}-2&=b_1^{-1} \left(C_{12}^{\mu,\rmh} + 4b_2/b_1\right)\,.
\end{align}
Combining equations~\eqref{eq:S3bias}~and~\eqref{eq:C12bias} allows us in principle to  solve for  the bias parameters,  by relying on theoretical  predictions for the dark matter cumulants on the one hand,
 and measurements for the halo cumulants on the other hand\footnote{These expressions closely resemble those given in \cite{BelMarinoni12} which use  a forward biasing model in the densities. This paper relies on the lowest order cumulants  predicted by tree-order perturbation theory and combines them in a difference that is suspected to be more robust than the individual cumulants.}.
 
This paper  extends this cumulant based strategy by taking advantage of the full two-point information \citep{BernardeauSchaeffer92,Munshi00}
which  consistently include higher order cumulants  leading to  improved accuracy, as demonstrated in \cite{Codis2016DE,Uhlemann17Kaiser}. 
In effect, instead of being restricted to the lowest order cumulants, it makes simultaneous use of the one-point PDF and the two-point sphere bias function. 
Indeed, it can be shown that the two-point sphere bias' slope with respect to the density is sensitive to bias alone, hence the joint analysis of both counts breaks the degeneracy. Appendix~\ref{sec:PTdegen}  sketches a proof at the perturbative level.  

\subsection{Joint one- and two-point statistics: a worked example} 
\label{sec:twoptapplication}
Let us finally present a worked out fiducial experiment
 that allows to simultaneously obtain the dark matter variance, correlation function as well as the bias parameters from measurements of 
one-point halo PDF $P_{\rmh}(\rho_\rmh|\sigma_{\mu,\rmm},b_0,b_1,b_2)$ given a redshift $z$ and sphere radius $R$ and the two-point halo sphere bias $b_\rmh(\rho_\rmh|\xi_\rmm(r),\sigma_{\mu,\rmm},b_0,b_1,b_2)$ at a separation $r\geq2R$. 
In practice, sampling the joint likelihood for 5 parameters is computationally expensive and tricky because the joint PDF is noisy and the signal coming from the sphere bias rather small\footnote{Note also that the tracer PDF's boundaries depend on the bias parameters, which, combined with the fact that the 
LDS model is only accurate on a finite range of densities adds an extra layer of complexity to the likelihood exploration.}. 
Let us therefore resort here to a simpler fitting procedure  to illustrate the capability of the one- and two-point halo statistics for jointly constraining the dark matter variance and correlation along with the bias parameters. 
A data sample is derived  from the simulation by binning the halo densities and measuring a histogram for the PDF $P_{\rmh}$ in the range $\rho_\rmh\in [0.1,3]$ with bin width $\Delta\rho_{\rmh,\mP}=0.01$ and the scaled halo sphere bias $\tilde b_{\circ,\rmh}$
\begin{equation}
\tilde b_{\circ,\rmh}(\rho_{\rmh})\equiv\left\langle\rho_{\rmh}'(r)|\rho_{\rmh}\right\rangle-1=\xi_{\circ,\rmh}b_{\circ,\rmh}(\rho_{\rmh}),
\end{equation}
 in the range $\rho_\rmh\in[0.07,2.5]$ with bin width $\Delta\rho_{\rmh,b}=3/21$. 
 The scaled halo sphere bias is used  instead of the halo sphere bias as this is the direct observable.
 The LDS prediction is given by 
  \begin{equation}
\tilde b_{\circ,\rmh}(\rho_{\rmh})=\left\langle \rho_{\rmh} b_{\circ,\rmm}(\rho_{\rmm}(\rho_{\rmh}))\right\rangle\,\, \xi_{\circ,\rmm} b_{\circ,\rmm}(\rho_{\rmm}(\rho_{\rmh})) \,,
\end{equation}
where the prefactor encodes the difference of the correlation function $\sqrt{\xi_{\circ,\rmh}/\xi_{\circ,\rmm}}=\langle \rho_{\rmh} b_{\circ,\rmm}(\rho_{\rmm}(\rho_{\rmh}))\rangle$ and is tabulated using a 5th order Taylor expansion of $\rho_{\rmh}(\rho_{\rmm})$ near one.
 
Using this  sample, a nonlinear model fit is implemented for the two functions $\mP_{\rmh}(\rho_\rmh)$ and $b_{\circ,\rmh}(\rho_{\rmh})$ with weights determined by the  errors from the measured PDF and bias function (using bootstrapping over  8 subsamples of the simulation). The result of the fit for the parameters and the associated uncertainties is given in Table~\ref{tab:jointfit} (see also Figure~\ref{fig:FoM-full} for the corresponding figures of merit) and agrees very well with the directly measured values reported in Table~\ref{tab:biasfit}. In particular, the sphere bias (i.e the two-point statistics of density in spheres) is shown as anticipated to break the degeneracy. 
 Since the dark matter correlation function $\xi_{\circ,\rmm}$ enters as an overall amplitude, the degeneracy is   broken by the information contained in the shape of the sphere bias function, rather than its amplitude, as can be seen perturbatively in Appendix~\ref{sec:PTdegen}.
As the noise is  more important in the two-point sphere bias than in the one-point density PDF, the error budget on the parameters of the model is dominated by the accuracy on the measurement of $\tilde b_{\circ,\rmh}$.

The total number of spheres  ($\approx 10^{6}$) is of the order of the number of spheres that a survey like Euclid will probe at a redshift around $z\approx 1$ \citep{Codis2016DE}. Hence one can expect this novel idea to be applicable to real data in a very near future, which will allow us to measure consistently the growth of fluctuations across cosmic time (through the dark matter variance $\sigma_{\mu,\rmm}$) and to characterise galaxy biasing (through a set of bias parameters at different redshifts). The accuracy of the constraints on those parameters depends on the accessible survey volume and therefore the number of spheres $N$, in a way which can be studied by subsampling the simulation. Redoing the above-described analysis on 8 subcubes of the simulation, yields
the average best fit values 
(notably 0.29, 0.030, 1.518, 0.181, 0.161 for $\sigma_{\mu,\rmm}$, $b_{0}$, $\beta_{1}$, $\beta_{2}$, $\eta$) are consistent with the parameters estimated from the full box (0.29, 0.032, 1.514, 0.177, 0.165), as seen on Figure~\ref{fig:subsampling}. The mean standard deviation 
are respectively 0.033, 0.0030, 0.0052, 0.0056, 0.0023 (to be compared with the one-sigma error bars from the full volume: 0.016, 0.00089, 0.0021, 0.0023, 0.00099),
which is consistent with a $1/\sqrt N$ scaling.
Overall, the typical one-sigma errors evolve as $\Delta \sigma_{\mu,\rmm} =0.016 \sqrt{10^6/N}$ and $\Delta \xi_{\circ,\rmm} =0.0017 \sqrt{10^6/N}$. 

The above presented  experiment is of course fairly idealized at various levels. It may turn out to be too optimistic, 
but should nonetheless provide a framework in which to implement a dark energy experiment based on count-in-cells.

\begin{table}
\centering
\label{tab:jointfit}
\begin{tabular}{| ll | cc | ccc}
\multicolumn{2}{c|}{param} & \multicolumn{2}{c|}{dark matter} & \multicolumn{3}{c|}{tracer bias}  \\\hline
$z$ & $R$ &  $\sigma_{\mu,\rmm} $& $\xi_{\circ,\rmm}(r) $& $b_{0}$ & $ b_{1}$ & $ b_{2}$\\\hline
\multirow{2}{*}{1} & \multirow{2}{*}{15} & $0.306$ & $0.0154$ & $0.0309$ & $0.463$ & $0.0534$\\
& & $\pm 0.015$ & $\pm 0.0016$ & $\pm 0.0016$ & $\pm 0.024$ & $\pm 0.0032$\\
\end{tabular}
\caption{Collection of the results (best fits and one-sigma confidence intervals) of the joint fitting procedure for $\mP_\rmh$ and $b_{\circ,\rmh}$ for radius $R$ [Mpc$/h$] and redshift $z$ at separation $r=30$Mpc$/h$. The expected values, as given in Table~\ref{tab:biasfit}, are $\sigma_{\mu,\rmm}=0.310$, $\xi_{\circ,\rmm}=0.016$, $b_0=0.028$, $b_1=0.473$, $b_2=0.055$ and lie well within the confidence intervals.}
\end{table}

\section{Conclusions}
\label{sec:Conclusion}
Starting from a very accurate model for the dark matter density-in-cells,  
we extended it to  biased tracers such as dark haloes or galaxies and compared them to the state-of-the-art Nbody simulation Horizon Run 4 in real and redshift space. 
Our main findings can be summarised as follows:
\begin{enumerate}
\item on scales of the order of 10 Mpc$/h$, mass-weighted subhalo densities show considerably less scatter than their number-weighted version; they can be accurately fit with a quadratic bias model in the log-densities and closely resemble the bias relation of mass-weighted galaxy densities.
\item Using a quadratic mean bias model for log-densities and neglecting the scatter is sufficient to obtain a one-point halo PDF and two-point sphere bias that are as accurate as the underlying dark matter results when compared against simulations, see Figures~\ref{fig:PDF_MEASURE_masslogbiasres}~and~\ref{fig:2ptbiashalo}. Combining the quadratic bias model with fitted coefficients with the dark matter PDF from large deviation statistics with the measured dark matter variance, the accuracy of the halo PDF is well within 5\% over a wide range of densities, in both real and redshift space.
\item The one-point PDF yields access to a one dimensional manifold in the four dimensional parameter space of dark matter variance and quadratic bias. 
\item Combining the one-point halo PDF and the two-point halo sphere bias, one can jointly constrain the nonlinear dark matter variance and correlation as well as the bias parameters, and hence disentangle tracer bias from nonlinear gravitational evolution.  This is of interest both from the point of view of dark energy  and non-linear power spectra estimation. {The density-dependent clustering signal encoded in the two-point sphere bias is related to the concept of `sliced' or `marked' correlation functions \citep[see e.g.][]{Sheth05,WhitePadmanabhan09,Neyrinck16}  which hence might contain valuable information about bias and could be used to break the degeneracy between linear bias and the clustering amplitude in the two-point correlation.}
\item Comparison to counts extracted from `full-physics' hydrodynamical simulations suggest that our findings will scale from dark halos to galaxies. 
\end{enumerate}

The excellent accuracy of the analytical prediction for the dark matter PDF and two-point bias plays a critical role in 
disentangling the dark matter variance from biasing when  applied to  tracers. Hence, this formalism should be applied  to constrain cosmology  using counts-in-cells statistics  in ongoing or upcoming surveys like DES, Euclid, WFIRST, LSST, KiDs, following the fiducial dark energy experiment presented in \cite{Codis2016DE}. 

\section*{Acknowledgements}
This work is partially supported by the grants ANR-12-BS05-0002 and  ANR-13-BS05-0005 of the French {\sl Agence Nationale de la Recherche}.
CU is supported by the Delta-ITP consortium, a program of the Netherlands organisation for scientific research (NWO)  funded by the Dutch Ministry of Education, Culture and Science (OCW). We thank Tobias Baldauf, Karim Benabed, Donghui Jeong, Marcello Musso, Fabian Schmidt, Ravi Sheth and the participants of the workshops `Statistics of Extrema in Large Scale Structure' and the `Biased Tracers of Large-Scale Structure' for discussions. We thank Iary Davidzon for having run the SED-fitting on the photometry of the simulated galaxies in the {\sc Horizon-AGN} simulation in order to compute mock observed stellar masses.
CU thanks IAP and CITA, while
MF, DP and SC also thank KIAS  for hospitality while some of this work was done.
Many thanks to St\'ephane Rouberol for smoothly running the Horizon cluster which is hosted by the Institut d'Astrophysique de Paris,
and to our colleagues who produced and post processed  the Horizon-AGN/ HR4 simulations.


\bibliographystyle{mnras}
\bibliography{DHbias}



\appendix

\section{Lognormal reconstruction}
\label{sec:LogNormal}
\begin{figure}
\centering
\includegraphics[width=1\columnwidth]{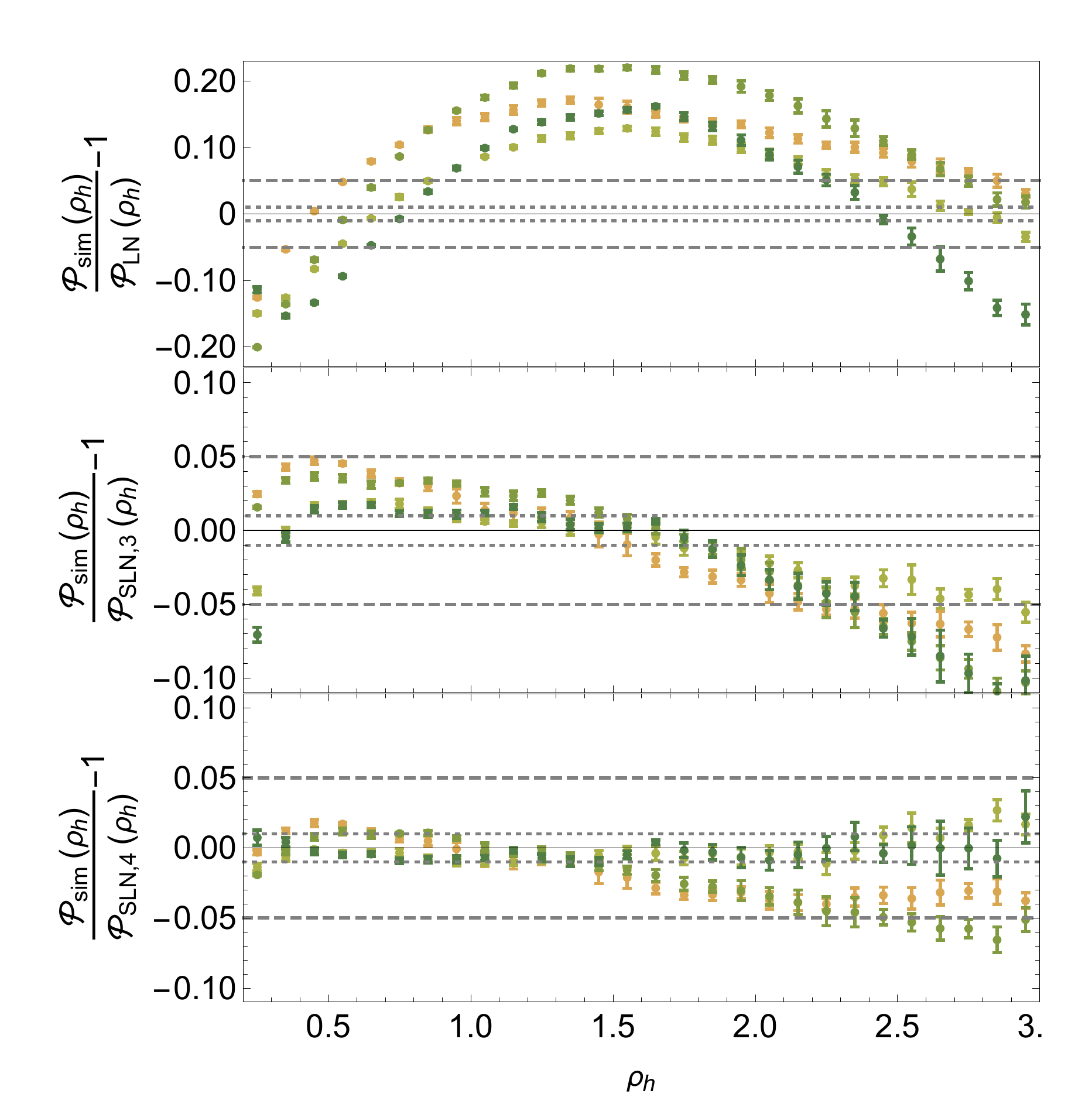}\\
\includegraphics[width=.92\columnwidth]{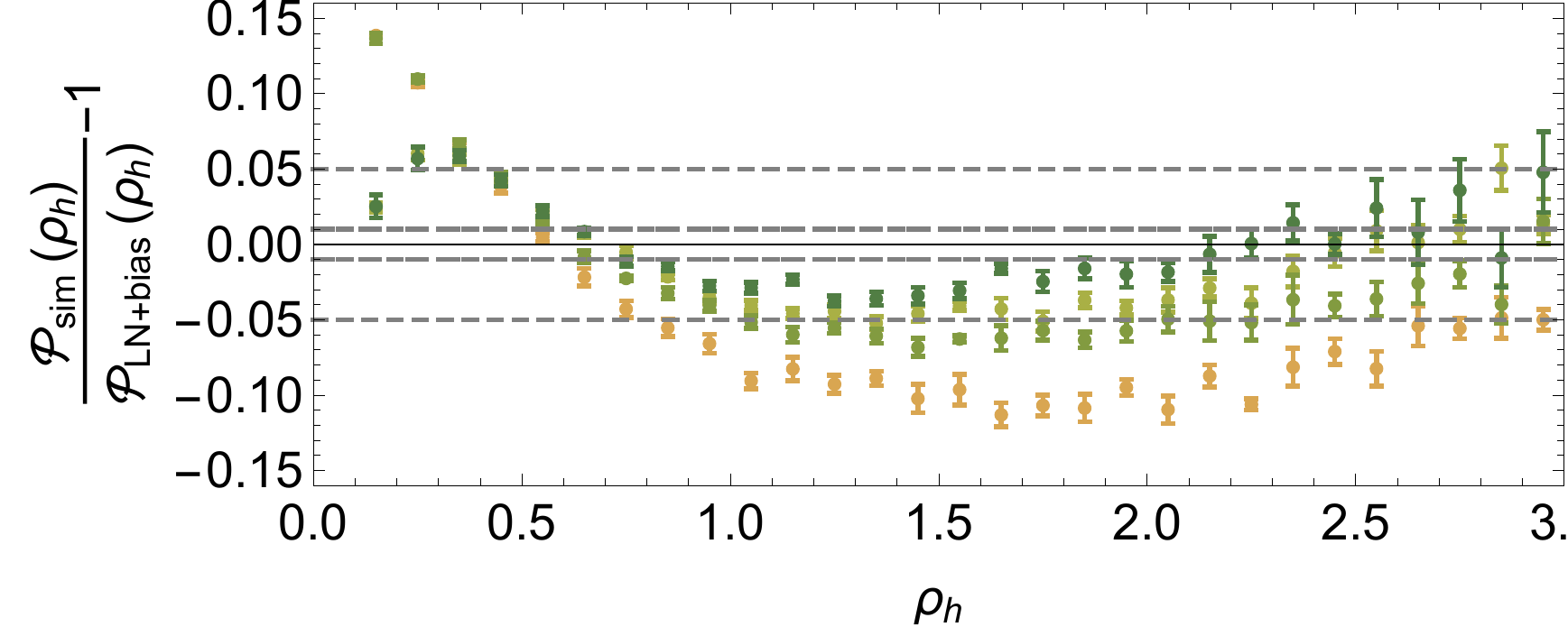}\\
\caption{Halo mass-density PDFs $\mP_{\rmh}$ for direct measurements based on halo catalogues at redshifts $z=0,1$ for radii $R=10, 15$ Mpc$/h$ in comparison to the recovered PDF of lognormal models with measured mean and variance (upper panel), the skewed lognormal model with measured cumulants up to skewness (upper middle panel) and up to kurtosis (lower middle panel) and the PDF assuming a lognormal model with measured matter variance and quadratic bias model for the log-densities (lower panel).
}
\label{fig:PDF_MEASURE_lognormalres}
\end{figure}

Let us  compare the LDS approach to the well-known lognormal models.
The lognormal PDF, proposed first from a dynamical model for dark matter in \cite{ColesJones91} but nowadays being used as a phenomenological parametrisation for PDFs of dark matter and its tracers, has the following form
\begin{align}
\label{eq:lognormal}
\mP_{\rm LN}(\rho\,|\,\sigma_\mu,\bar\mu) &=\frac{1}{\sqrt{2\pi\sigma_\mu}} \frac{1}{\rho} \exp\left[-\frac{(\log\rho -\bar\mu)^2}{2 \sigma_\mu^2}\right]\,,
\end{align} 
where the variance $\sigma_\mu$ of the log-density $\mu=\log\rho$ can be treated as free parameter and the mean of the log-density is connected to the variance via $\bar\mu=\langle\log\rho\rangle\approx-\sigma^2/2$ by requiring a unit mean density $\langle\rho\rangle=1$. 
The skewed lognormal PDF as introduced in \cite{Colombi1994},  involves an Edgeworth expansion around the lognormal PDF and reads
\begin{align}
\label{eq:skewedlognormal}
\mP_{\rm SLN}(\rho\,|\,\sigma_\mu,\bar\mu,\epsilon_3,\epsilon_4) &= \mP_{\rm LN}(\rho\,|\,\sigma_\mu,\bar\mu) \\
\notag & \hskip -1cm\times \left[1+\frac{\epsilon_3}{6} H_3(\hat{\mu}) + \frac{\epsilon_4}{24}H_4(\hat{\mu}) + \frac{\epsilon_3^2}{72}H_6(\hat{\mu})\right] \,.
\end{align}
with the normalised log-density $\hat\mu=(\mu-\bar\mu)/\sigma_\mu$, its rescaled cumulants $\epsilon_n=\langle \hat\mu\rangle_c$ and the probabilist's Hermite polynomials $H_n$\footnote{Note that Edgeworth expansions are known to be problematic in the tails of the distribution because the expression in the brackets can eventually become negative depending on the size of the corrections in the cumulant expansion.}. A comparison between the accuracy of the three different lognormal based models, when the underlying parameters (the mean $\bar\mu$, variance $\sigma_\mu$, skewness $\epsilon_3$ and kurtosis $\epsilon_4$ of the log-density) are measured from the simulated halo densities is shown in Figure~\ref{fig:PDF_MEASURE_lognormalres}. The generalized normal distribution $\mathcal N_{v2}$ adopted by \cite{Shin17} to fit dark matter PDFs has very similar properties to the skewed lognormal PDF in the range of radii we consider and hence will not be discussed here.

The lognormal dark matter PDF can be combined with a polynomial bias model for the log-densities. 
For a linear bias model of the log-densities, the resulting halo PDF is again lognormal with variance and mean given by 
$\sigma_{\mu,\rmh}=\sigma_{\mu,\rmm}/b_1$ and $-\bar\mu_\rmh=b_0/b_1 + b_1 \sigma_{\mu,\rmh}^2/2$, once the dark matter mean density is fixed to one so that $-\bar\mu_\rmm=\sigma_{\mu,\rmm}^2/2$. In addition, the halo mean density being one, one gets an additional contraint which relates the constant bias shift to the linear bias factor and the variance according to $b_0=b_1(1-b_1)\sigma_{\mu,\rmh}^2/2$
and agrees with the leading order perturbative result. In this model, there is a full degeneracy between the linear bias $b_1$ and the log-variance of the underlying dark matter $\sigma_{\mu,\rmm}$.

Let us now consider a quadratic log-bias model. Even if the dark matter PDF was close to lognormal  \citep[which is typically the case at $\sim10\%$ accuracy, see][]{Uhlemann17Kaiser}, this nonlinear mapping induces extra terms in the exponential. If one expands these terms in an Edgeworth-like fashion, one can see that nonlinear bias naturally feeds into higher order cumulants, in particular the skewness and kurtosis, which is why it is necessary to go to the skewed lognormal forms to fit the measured halo PDF. {The residuals obtained when augmenting the lognormal dark matter PDF with the quadratic bias model with measured parameters are shown as comparison in the lower panel of Figure~\ref{fig:PDF_MEASURE_lognormalres}.}
{In this case, the predicted PDF is slightly less accurate than the large-deviation prediction, with two additional parameters that cannot easily be related to bias because they mix in contributions from the dark matter PDF, which is significantly better fitted by a skewed lognormal. This is in contrast with the LDS formalism that clearly disentangles the effect of gravitational evolution (parametrised through the nonlinear dark matter variance) from  nonlinear biasing (parametrised through the bias parameters).}

\section{Breaking degeneracies}
\label{sec:PTdegen}
 The main text has shown  that with a practical implementation of the LDS formalism, one can accurately measure bias parameters and dark matter variance,
  and break the degeneracies by including information from  two-point statistics, an idea also followed by \cite{BelMarinoni12} in another context.
Let us illustrate these findings using perturbation theory.

\subsection{One-point PDF}
From equation~(\ref{eq:POLYBIASloginv}), one can easily compute the relation between halo and matter contrast within the quadratic log bias model
\begin{equation}
\delta_{\rmh}=-1+\exp\left[\frac{\sqrt{b_{1}^{2}-4b_{2}(b_{0}-\log(1+\delta_{\rmm}))}-b_{1}}{2b_{2}}\right].
\end{equation}
Expanding this relation for small contrasts yields perturbative bias consistency relations.
First, imposing a zero mean for the halo contrast allows to get $b_{0}$ at all orders in the dark matter variance $\sigma$
\begin{equation}
b_{0}=\textstyle \sum_{i} b_{0}^{(i)}\sigma^{2i} \,,
\end{equation}
with
\begin{equation}
\label{eq:b0ofb1b2}
b_{0}^{(0)}=0\,, \quad
b_{0}^{(1)}=\frac{b_{1}-2b_{2}-b_{1}^{2}}{2b_{1}^{2}}.
\end{equation}
The measured halo variance then imposes a relation between the dark matter variance and the bias parameters which reads 
\begin{equation}
\label{eq:b1deg}
\sigma_\rmh^{2}=\left(\frac{\sigma}{b_{1}}\right)^{2}\left[1+\sigma^{2}\Delta_{\rm NL}\right]\,,
\end{equation}
with
\begin{equation}
\Delta_{\rm NL}^{(0)}=\frac{(S_{3}-3)(b_{1}-2b_{2}-b_{1}^{2})}{b_{1}^{2}}+\frac{ 20b_{2}^{2}-8b_{1}b_{2}-b_{1}^{4}}{2
   b_{1}^{4}}.
\end{equation}
The constraints are therefore dominated by this degeneracy between $b_{1}$ and $\sigma$ at first order in PT.
After the mean and the variance, the PDF will typically pick up the information from the skewness. Let us therefore compute perturbatively the skewness of the halo density field. At first order, it reads
\begin{equation}
S_{3,\rmh}=3+b_{1}(S_{3}-3)-6\frac{b_{2}}{b_{1}}+{\cal O}(\sigma).
\end{equation}
This latter equation gives a relation between $\sigma$ and $b_{2}$ at first order
\begin{equation}
\label{eq:b2deg}
b_{2}=\frac{\sigma (3-S_{3,\rmh})}{6\sigma_\rmh}\left(1 +\frac{\sigma(S_{3}-3)}{\sigma_\rmh(S_{3,\rmh}-3)}\right)+{\cal O}(\sigma^{2}).
\end{equation}
Equations~\eqref{eq:b1deg} and \eqref{eq:b2deg} predict a linear degeneracy between on the one hand $\sigma$ and $b_{1}$ and on the other hand $\sigma$ and $b_{2}$ which is indeed observed when performing the model fitting (see Figure~\ref{fig:dline}).
This model fitting described in Section~\ref{sec:Applications} eventually gathers all the information coming from the mean, variance, skewness and higher order cumulants in a fully consistent way (because LPD provides the PDF and therefore the full statistics). 
In principle the knowledge of the full hierarchy of cumulants eventually break those degeneracies if the LDS model if exact. In practice,
i) sample noise prevents accurate measurements of the higher order cumulants (kurtosis etc) which scale like higher power of the variance ($\sigma^{6}$ and above);
 ii) loop corrections in the skewness that are not accounted for in the LDS model appears at the same perturbative order as those higher order cumulants and therefore do not allow us to fully break the degeneracy between the parameters. 
To break this degeneracy, one must involve  two-point statistics as described in the next section.

\subsection{Two-point PDF}
\label{app:2ptbias}
Let us assume that the two-point PDF of the matter density is well described by its large-scale approximation given by equation~(\ref{eq:fullPDFlargeseparation}).
The sphere bias $b_{\circ,\rmm}(\rho_{\rmm})$ can be exactly computed using the large-deviation principle \citep{Codis2016twopoint,Uhlemann17Kaiser}. A fair approximation for small densities is given by
\begin{equation}
 b_{\circ,\rmm}(\rho_{\rmm})=\frac{\tau_{\rm SC}(\rho_{\rmm})}{\sigma_{L}^{2}(R\rho_{\rmm}^{1/3})}.
\end{equation}
Remarkably,  plugging  in the bias relation in $b_{\circ}(\rho)$, shows that the sphere bias of the halo density field behave at small density as
\begin{equation}
b_{\circ,\rmh}(\rho_{\rmh})=\frac{\delta_{\rmh}b_{\circ}^{(1)}+b_{\circ}^{(0)}}{  \sigma_{L}^{2} \left(R\sqrt[3]{e^{b_0}+1}\right)}\,,
\end{equation}
where
\begin{align}
b_{\circ}^{(1)}&\!= \!\frac{2
   e^{b_0} b_1 \nu  \gamma\left(\!\sqrt[3]{e^{b_0}+1}\right)}{3 \left(e^{b_0}+1\right)^{2/3}
   }\!\!
   \left[\left(e^{b_0}\!+1\right)^{-\frac{1}{\nu} }\!-\!1\right]\!+\frac{
   e^{b_0} b_1}{\left(e^{b_0}+1\right)^{\frac{1+\nu}{\nu} }}\,,\nonumber
   \\
   b_{\circ}^{(0)}&=\nu 
\left(1-\left(e^{b_0}+1\right)^{-\frac{1}{\nu} }\nonumber
\right)\,. \end{align}
and $\gamma=\sigma'/\sigma$.
Obviously, the overall amplitude in $b_{\circ,\rmh}$ cannot be measured because it is degenerate with the unknown dark matter correlation function $\xi_{\circ,\rmm}$ but the ratio between the slope called $b_{\circ}^{(1)}$ and intercept $b_{\circ}^{(0)}$ can
\begin{equation}
\frac{b_{\circ}^{(1)}}{b_{\circ}^{(0)}}\!=\!- \frac{2
   e^{b_0} b_1  \gamma\left(\!\sqrt[3]{e^{b_0}+1}\right)}{3 \left(e^{b_0}+1\right)^{2/3}
   }
 +\frac{
   e^{b_0} b_1\left(e^{b_0}+1\right)^{-1 }
}{
\nu 
\left(\left(e^{b_0}+1\right)^{\frac{1}{\nu} }-1\right)
}. \label{eq:breakb1}
\end{equation}
This ratio is in particular proportional to $b_{1}$ and does not depend on the variance. 
Constraining this ratio, as is done in the joint fit presented in the main text
will therefore break the degeneracy  in equation~\eqref{eq:b1deg}.

\balance

\section{The Horizon-AGN simulation}
\label{sec:hagn}

\begin{figure}
\centering
\includegraphics[width=1\columnwidth]{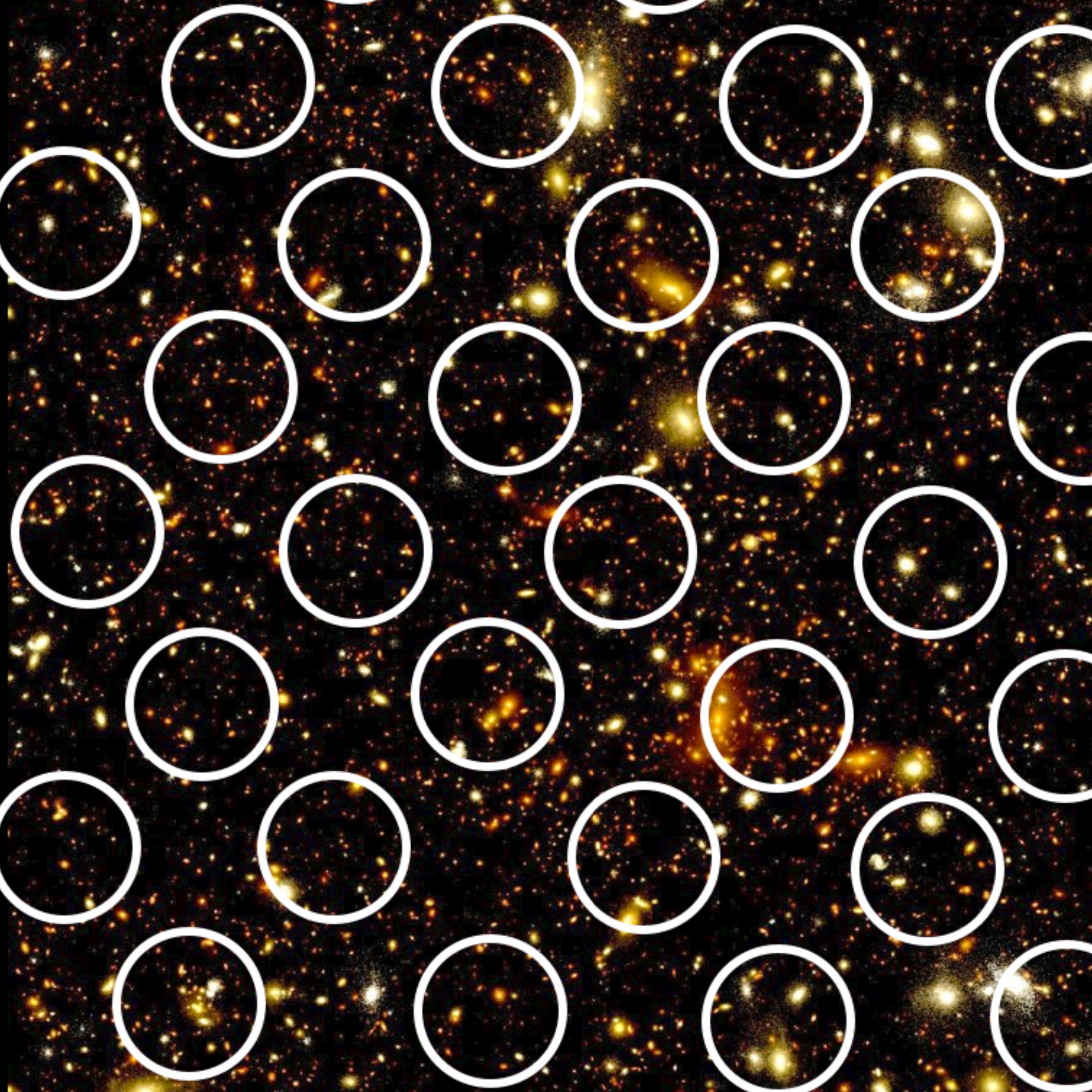}\\
\caption{Qualitative distribution of spheres in the simulated galaxies in Horizon-AGN.
The background represents synthetic galaxies produced by the simulation while converting cold gas into stars.
Realistic colours are post processed using spectral synthesis \citep{kaviraj17}.}
\label{fig:horizon-AGN}
\end{figure}

Let us briefly describe the cosmological hydrodynamical simulation used in the main text, horizon-AGN \citep{duboisetal14}.
The  simulation {\url{http://www.horizon-simulation.org/}} is run with a $\Lambda$CDM cosmology with total matter density $\Omega_{\rm  m}=0.272$, dark energy density $\Omega_\Lambda=0.728$, amplitude of the matter power spectrum $\sigma_8=0.81$, baryon density $\Omega_{\rm  b}=0.045$, Hubble constant $H_0=70.4 \, \rm km\,s^{-1}\,Mpc^{-1}$, and $n_s=0.967$ compatible with the WMAP-7 data~\citep{komatsuetal11}.
The size of the simulation box is $L_{\rm  box}=100 \, h^{-1}\rm\,Mpc$ on a side, and the volume contains $1024^3$ DM particles, corresponding to a DM mass resolution of $M_{\rm  DM, res}=8\times 10^7 \, \rm M_\odot$.
The simulation is run with the {\sc ramses} code~\citep{teyssier02}, and the initially coarse $1024^3$ grid is adaptively refined down to $\Delta x=1$ proper kpc, with refinement triggered in a quasi-Lagrangian manner: if the number of DM particles becomes greater than 8, or the total baryonic mass reaches 8 times the initial DM mass resolution in a cell.
It lead to a typical number of $6.5\times 10^9$ gas resolution elements (leaf cells) in the simulation at $z=1$.
Heating of the gas from a uniform UV background takes place after redshift $z_{\rm  reion} = 10$ following~\cite{haardt&madau96}. 

Star formation occurs in regions of gas number density above $n_0=0.1\, \rm H\, cm^{-3}$ following a Schmidt law: $\dot \rho_*= \epsilon_* {\rho_{\rm g} / t_{\rm  ff}}$,  where $\dot \rho_*$ is the star formation rate mass density, $\rho_{\rm g}$ the gas mass density, $\epsilon_*=0.02$ the constant star formation efficiency, and $t_{\rm  ff}$ the local free-fall time of the gas.
Feedback from stellar winds, supernovae type Ia and type II are included into the simulation with mass, energy and metal release.
The simulation also follow the formation of black holes (BHs), which can grow by gas accretion at a Bondi-capped-at-Eddington rate and coalesce when they form a tight enough binary.
BHs release energy in a quasar/radio (heating/jet) mode when the accretion rate is respectively above and below one per cent of Eddington, with efficiencies tuned to match the BH-galaxy scaling relations at $z=0$~\citep[see][for details]{duboisetal12agnmodel}.
A lightcone has been generated from the simulation,  as described in \cite{pichon2010}. The area of the lightcone is 5 deg$^{2}$ below $z=1$, and 1 deg$^{2}$ above. A mock photometric galaxy catalog has been extracted from the lightcone in order to mimic observational datasets (see Laigle et al. in prep for more details). 
Galaxies have been identified from the stellar particles distribution using the {\sc AdaptaHOP} halo finder \citep{aubert04}. The local density is computed from a total of 20 neighbours, and a density threshold $\rho_{\rm t}$ of 178 times the average matter density is required to select structures. Once identified mock galaxies in the lightcone, a BC03 simple stellar population (SSP) has been attached to any stellar particle in each galaxy, according to its mass and stellar metallicity. The spectrum of the galaxy is then obtained by adding the SEDs of all the SSPs. The (possibly redshifted) spectra are then convolved with photometric filter passbands, in order to get absolute and apparent magnitudes in the following 13 bands: $NUV$, $u$, $B$, $V$, $r$, $i^{+}$, $z^{++}$, $Y$, $J$, $H$, $K_{\rm s}$, $3.6\mu$m, $4.5\mu$m. Dust attenuation is also taken into account along the line of sight of each stellar particle in the galaxy, assuming the dust mass  scales with the gas metal mass, with a dust-to-metal ratio of 0.4 \citep{dwek1998,jonsson2006}. 
In order to get observed stellar masses, the SED-fitting code {\sc LePhare} \citep{arnouts02,ilbert06} has been run using as input photometry the virtual magnitudes included in the mock catalogue and with a configuration similar to  \citet{laigle2016}. 

In closing the galaxy population was shown to reproduce in overall the luminosity function of observed galaxies in \cite{kaviraj17} (see Figure~\ref{fig:horizon-AGN} for a qualitative representation of the count-in-cells within its lightcone). 

\end{document}